\UseRawInputEncoding
\documentclass[prx,twocolumn,aps]{revtex4-2}
\usepackage{graphicx}
\usepackage[dvipsnames]{xcolor}
\usepackage{dcolumn}
\usepackage{bm}
\usepackage{amsmath}
\usepackage{amssymb}
\usepackage{amsthm}
\usepackage{cancel}
\usepackage{chemarr}
\usepackage{centernot}
\usepackage{hyperref}
\usepackage{mhchem}
\usepackage{booktabs}

\usepackage[english]{babel} 

\definecolor{webgreen}{rgb}{0,.5,0}
\definecolor{webbrown}{rgb}{.6,0,0}
\definecolor{grigio}{rgb}{.85,.85,.85} 
\definecolor{RoyalBlue}{rgb}{0.0, 0.14, 0.4}
\definecolor{skyblue1}{rgb}{0.45,0.62,0.81}
\definecolor{skyblue2}{rgb}{0.2,0.39,0.64}
\definecolor{skyblue3}{rgb}{0.13,0.29,0.53}
\definecolor{scarlet1}{rgb}{0.93,0.16,0.16}
\definecolor{scarlet2}{rgb}{0.8,0,0}
\definecolor{scarlet3}{rgb}{0.64,0,0}
\definecolor{g}{gray}{0.50}
\hypersetup{%
    colorlinks=true, linktocpage=true, pdfstartpage=1, pdfstartview=FitV,%
    breaklinks=true, pdfpagemode=UseNone, pageanchor=true, pdfpagemode=UseOutlines,%
    plainpages=false, bookmarksnumbered, bookmarksopen=true, bookmarksopenlevel=1,%
    hypertexnames=true, pdfhighlight=/O,
    urlcolor=webbrown, linkcolor=RoyalBlue, citecolor=webgreen, 
    pdfsubject={},%
    pdfkeywords={},%
    pdfcreator={pdfLaTeX},%
    pdfproducer={LaTeX REVTeX}%
    }
\begin{document}

\title{Energy Transduction in Complex Networks with Multiple Resources:\\ The Chemistry Paradigm}
\author{Massimo Bilancioni}
\email{massimo.bilancioni@uni.lu}
\affiliation{Complex Systems and Statistical Mechanics, Department of Physics and Materials Science, University of Luxembourg, 30 Avenue des Hauts-Fourneaux, L-4362 Esch-sur-Alzette, Luxembourg}
\author{Massimiliano Esposito}
\email{massimiliano.esposito@uni.lu}
\affiliation{Complex Systems and Statistical Mechanics, Department of Physics and Materials Science, University of Luxembourg, 30 Avenue des Hauts-Fourneaux, L-4362 Esch-sur-Alzette, Luxembourg}

\date{\today}

\begin{abstract}
We extend the traditional framework of steady state energy transduction---typically characterized by a single input and output---to multi-resource transduction in open chemical reaction networks (CRNs). Transduction occurs when stoichiometrically balanced processes are driven against their spontaneous directions by coupling them with thermodynamically favorable ones.
However, when multiple processes (resources) interact through a shared CRN, identifying the relevant set of processes for analyzing transduction becomes a critical and complex challenge. To address this, we introduce a systematic procedure based on elementary processes, which cannot be further decomposed into subprocesses.
Our theory generalizes the methodology used to define transduction efficiency in thermal engines operating between multiple heat baths. By selecting a reference equilibrium environment, it explicitly reveals the inherently relative nature of transduction efficiency and ties its definition to exergy. This framework also allows one to exclude unusable outputs from efficiency calculations.
We further extend the concept of chemical gears to multi-process transduction, demonstrating their versatility as an analytical tool in complex settings. Finally, we apply our framework to central metabolic pathways, uncovering deep insights into their operation and highlighting the crucial difference between thermodynamic efficiencies and stoichiometric yields.
\end{abstract}

\maketitle

\section*{Introduction}

Free energy transduction is a fundamental process underlying a wide range of phenomena in both natural and engineered systems. While theoretical frameworks have mostly focused on energy transduction with a single input and output~\cite{HILL_Free_En_Transd_Bio_1977, Parmeggiani_E_transduction_isoT_ratchets_1999,  seifertStochasticThermodynamicsFluctuation2012, penocchioThermodynamicEfficiencyDissipative2019, Brown_Noneq_Free_Energy_Trans_2020, Large_Transduction_in_aut_system_2021}, many real-world systems deviate from this simplicity, performing transduction across multiple resources.
For example, cells metabolize a variety of substrates---such as glucose, proteins, and fatty acids---to extract free energy, which drives essential processes like biosynthesis, transport, and signaling. Similarly, in food webs, organisms derive energy from diverse resources---ranging from specific prey to plant matter---and allocate it to essential functions such as metabolism, growth, reproduction, and ecological interactions. In engineered systems, power grids integrate energy inputs from fossil fuels, nuclear power, and renewables, distributing it to industrial operations, residential applications, transportation, and energy storage systems. Defining efficiency in these systems requires a rigorous framework to address conceptual challenges---such as identifying inputs and outputs, accounting for maintenance costs, recycling, and downstream losses---while ensuring consistency across studies, avoiding case-specific, non-generalizable definitions.

In this paper, we present a comprehensive framework for analyzing energy transduction in open chemical reaction networks (CRNs), focusing on stationary conditions. 
Open CRNs provide an ideal platform for studying multi-resource transduction in complex networks for several key reasons.
First, they are ubiquitous in nature, with examples spanning different fields and scales, including biochemistry~\cite{Voet_Fundamentals_of_Biochemistry_2013}, combustion~\cite{Gardiner_Combustion_Chemistry_1984}, ecology~\cite{Garvey_Trophic_Ecology_2016}, geochemistry~\cite{McSween_GEOchem_Pathway_processes_2003}, biogeochemistry~\cite{Schlesinger_Biogeochemistry_2013,Smith_Fourth_geosphere_2016},
atmospheric chemistry~\cite{Warneck_Chemistry_Natural_Atmosphere_1999, Wayne_Chemistry_Atmosphere_2006}, and astrochemistry~\cite{Williams2023Astrochemistry}.
Second, they are inherently complex because, unlike systems that process a single conserved quantity---such as energy in thermal engines or charge in electrical circuits---they function like a marketplace, where a large variety of molecular species exchange groups of atoms through reactions. These reactions naturally form a hypergraph, capturing the many-to-many relationships inherent in CRNs. 
Third, most CRNs exchange a multitude of molecular species with their environment. These species, known as \emph{chemostatted} species, are maintained at fixed concentrations over relevant time scales by their respective environmental reservoirs, called \emph{chemostats}, which set the CRN’s {\it operating conditions}.
As we will show, the transformation induced on the environment by the chemical reactions can be decomposed into  chemical processes, which are stoichiometrically balanced conversions among chemostatted species and can be viewed as effective reactions between them.
Transduction occurs when free energy is transferred between chemical processes---specifically, when exergonic processes, running thermodynamically downhill, drive endergonic processes uphill.
{\it Simple transduction} involves a single exergonic process coupled to a single endergonic one. However, in most cases, open CRNs perform {\it multi-resource transduction}, involving multiple interacting processes, which establishes them as paradigmatic multi-resource transducers.

The first systematic framework for analyzing transduction in CRNs was pioneered by Hill~\cite{HILL_Free_En_Transd_Bio_1977} but was restricted to (pseudo)linear CRNs---the simplest type of CRNs, where the hypergraph structure reduces to a simple graph. 
More recently, a theoretical framework was developed to extend transduction analysis to nonlinear CRNs~\cite{wachtelFreeenergyTransductionChemical2022}. 
While this framework effectively handles simple transduction, it overlooks a crucial issue in the multi-resource case: it offers no clear guidelines for selecting the processes in terms of which transduction is analyzed.
This selection is critical because there are numerous possible choices, each leading to distinct definitions of efficiency. Moreover, an incorrect selection can render transduction ambiguous or even undetectable.

The first part of this work addresses this gap by outlining the key requirements for analyzing transduction and establishing a systematic procedure for selecting the correct set of processes.
This procedure has two key conceptual highlights:
First, it generalizes the framework for defining transduction in thermal engines connected to multiple heat baths~\cite{Bejan2016AdvancedThermodynamics}. As in that context, a reference equilibrium environment is specified, which defines the chemical \emph{exergy}---that is, the maximum work extractable from out-of-equilibrium chemostats relative to that reference equilibrium. By adopting this approach, we adhere to a physically meaningful definition of efficiency, expressed as the ratio of output to input exergy.
Second, we analyze transduction in terms of \emph{elementary} processes---those that cannot be further decomposed into subprocesses---a new concept introduced in this work.
These processes represent the smallest fundamental units of free energy that can be produced or consumed in the chemostats. They are defined exclusively by the atomic composition of chemical species, using the atomic composition matrix as the primary analytical tool. 
By basing our analysis on elementary processes, we ensure that all transduction occurring in the CRN is captured. 
Once the set of processes is selected, we prescribe a method for classifying them as input or output processes, enabling a well-defined measure of transduction efficiency. Unlike previous studies, our framework allows for the exclusion of ‘unusable’ outputs, accounting for potential downstream losses that are not captured within the CRN.
Additionally, we emphasize the inherently relative nature of transduction efficiency, as it may depend on the choice of the reference equilibrium, the definition of ‘useful’ outputs, and the scope of the CRN analyzed.

In the second part of this paper, we build on our previous work~\cite{bilancioni2025gears}, where we demonstrated that, like most human-made machines, CRNs possess \emph{chemical gears}. These gears---representing multiple transduction pathways, each with a distinct efficiency---provide a refined understanding of the second law of thermodynamics. Specifically, they allow one to determine a CRN’s optimal efficiency and identify the specific gear responsible for achieving it as a function of the operating conditions. Remarkably, this analysis depends solely on the network's topology. However, our earlier framework was limited to CRNs undergoing simple transduction. Here, we extend these findings to multi-resource transduction, demonstrating that chemical gears remain a powerful analytical tool even in this more general setting. 

\begin{figure*}[htbp!]
    \centering
    \includegraphics[width=1.02\linewidth]{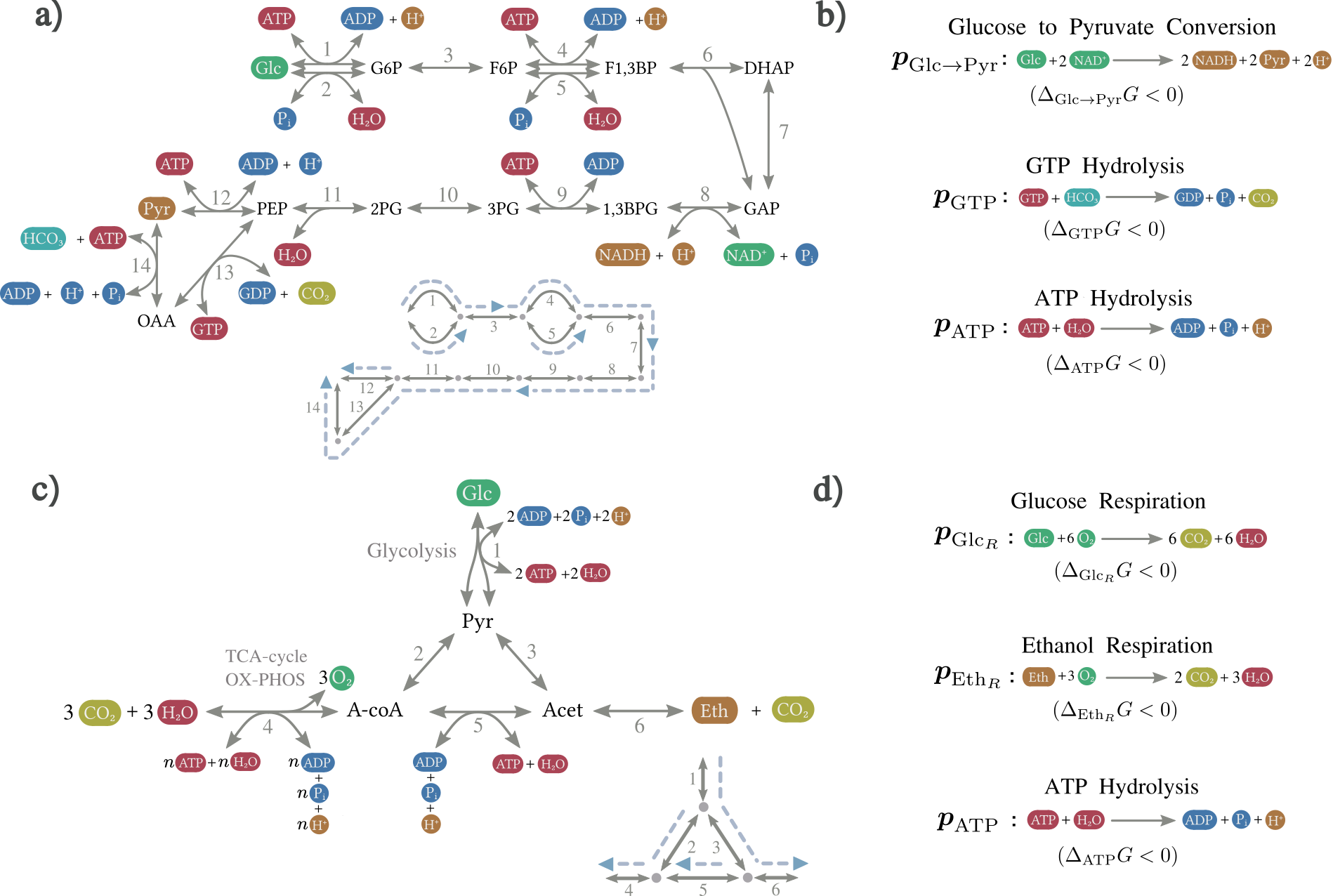}
    \caption{
    \textbf{a) + c)} Open CRN encompassing glycolysis and gluconeogenesis (GG network) alongside a schematic open CRN representing yeast's respiration and fermentation metabolism (RF network). Chemostatted species are highlighted. Below each respective CRN, a schematic representation emphasizes their structural organization, with dashed arrows indicating the forward direction (by convention) of reactions.
    \textbf{b) + d)} Sets of chemical processes through which transduction is analyzed in the corresponding networks. Throughout the paper, the sign imposed to the associated Gibbs free energy changes is chosen according to physiological conditions (Table~\ref{Tab:Standard physiological conditions}).
    }
    \label{fig:1}
\end{figure*}

Finally, in the last part of the paper, we demonstrate the practical relevance of our framework by applying it to multi-resource metabolic networks. 
We analyze a network encompassing glycolysis and gluconeogenesis pathways and a network representing the respiration and fermentation metabolism of yeasts (Fig.~\ref{fig:1}). For given operating conditions, we identify the gears available for transduction and the maximum efficiency accessible by the CRN. To do so, we rely solely on CRN stoichiometry without requiring any knowledge of the fluxes within the network. 
These analytical results---unavailable in previous transduction frameworks---provide a powerful benchmark for assessing, through experimental flux measurements or simulations, the extent to which flux regulations are contingent on thermodynamic efficiency considerations. 
For instance, this approach could be used to determine whether metabolic switching points---where an organism’s metabolic fluxes undergo significant rearrangement in response to changing operating conditions---are driven by efficiency or other considerations. 
Our analysis reveals intriguing findings. 
In the case of glycolysis and gluconeogenesis, experimental flux measurements indicate that, out of eight thermodynamically available gears, only the two least efficient ones are used to transduce in living organisms, raising the question of why this selection occurs.
For the second network, our analysis reveals that fermentation achieves higher thermodynamic efficiency than respiration under physiological conditions, despite its lower stoichiometric yield. This highlights the importance of distinguishing thermodynamic efficiency from stoichiometric yield, the latter being a non-thermodynamic metric that is commonly used~\cite{Molenaar_Shifts_growth_strategies_reflect_tradeoffs_2009,Flamholz_2013_Glycolytic_strategy_tradeoff,Pfeiffer_Evol_perspect_Crabtree_2014,Basan_Overflow_metabolism_Ecoli_proteome2015,Yang_Principles_proteome_allocation2016,Hu2023_Proteome_Efficiency}.\\

This paper is structured as follows. In Sect.~\ref{Sect:Setup} we introduce the setup, in Sect.~\ref{Sect:Cycles} we review the concept of CRN cycles, and in Sect.~\ref{Sect:Atomic composition matrix} we define the atomic composition matrix. Sect.~\ref{Sect:Chemical processes} formalizes the notions of chemical processes and introduces the subclass of elementary processes. 
Sect.~\ref{Sect:Complete and independent sets of processes} defines complete and independent sets of processes and their properties.
Sect.~\ref{Sect: Requirements for a Set of Processes for transduction} outlines the essential criteria that a set of processes must satisfy to be suitable for a transduction analysis. Sect.~\ref{Sect:Reference Equilibrium} shows how establishing a reference equilibrium  naturally determines a special set of processes and the exergies of chemical species. Building on these results, Sect.~\ref{Sect:Procedure for Selecting the Set of Processes} presents a systematic procedure to select an appropriate set of processes for transduction. In Sect.~\ref{Sect:Def of efficiency}, we define transduction efficiency, emphasizing its relative nature. Sect.~\ref{Sect:Gears} extends the concept of chemical gears to multi-process transduction: We start by reviewing the notion of elementary flux modes (EFMs) in Sect.~\ref{Sect:EFMs}, which are used in Sect.~\ref{Sect:Def gear efficiency} to define gears and their efficiencies. In Sect.~\ref{Sect:Conformal gears}, we introduce the concept of gear conformality. Sect.~\ref{Sect:Upper bound} presents the topological upper bound on transduction efficiency established by gears, with applications to the two illustrative metabolic networks in Sect.~\ref{Sect:Applications}.
Discussions and conclusions are laid out in Sect.~\ref{Sec:DiscConc}.

\section{Open Chemical Reaction Networks}

In this section, we review the central concepts needed in the rest of the paper to describe CRNs.

\subsection{Setup}
\label{Sect:Setup}

We consider open and well-stirred chemical reaction networks (CRNs) under stationary conditions, consisting of a set of reversible reactions $\rho$, internal species $X$, and external species $Y$. All reactions are mass- and charge-preserving conversions among species. The external species are chemostatted, meaning they are continuously exchanged with the environment, which ensures that their concentrations---and thus their chemical potentials---remain constant over time.
Figures~\ref{fig:1}a and \ref{fig:1}c illustrate the two representative open CRNs analyzed in this paper. The first depicts a network combining glycolysis and gluconeogenesis (GG network)~\cite{wachtelFreeenergyTransductionChemical2022}, while the second represents a coarse-grained yet thermodynamically complete model of yeast metabolism capturing both respiration and fermentation (RF network)~\cite{Pfeiffer_Evol_perspect_Crabtree_2014}. As detailed in Sects.~\ref{Sect:Atomic composition matrix} and \ref{Sect:Local validity of the Second law}, this simplified representation retains all necessary information for our analysis.
In both cases, the $Y$ species are highlighted.
The state of the system is defined by the vector of concentrations 
\begin{equation}
    \boldsymbol{z} = \begin{pmatrix} \boldsymbol{x} \\ \boldsymbol{y}  \end{pmatrix},
\end{equation}
where  $\boldsymbol{x} = (\dots, [x_i],\dots)$ denotes the concentrations of the internal species $X$ and  $\boldsymbol{y} = (\dots, [y_i],\dots)$  denotes the concentrations of the external species $Y$.
The evolution of these concentrations is governed by the stoichiometric matrix $\mathbb{S}$~\cite{polettiniIrreversibleThermodynamicsOpen2014a}, which encodes the topology of the CRN by specifying how reactions interconvert species. Defining this matrix requires establishing the forward direction for each reaction.
By separating the stoichiometric matrix with respect to internal and external species, 
\begin{equation}
    \mathbb{S} = \begin{pmatrix} \mathbb{S}^X \\ \mathbb{S}^Y \end{pmatrix},
\end{equation}
one can write the CRN's dynamics as 
\begin{subequations}
    \label{eq:crn_dynamics}
    \begin{equation}
        d_t{\boldsymbol{x}}=\mathbb{S}^X \boldsymbol{J},
    \end{equation}
    \begin{equation}
        d_t{\boldsymbol{y}}=\mathbb{S}^Y \boldsymbol{J}+\boldsymbol{I}^Y.
    \end{equation}
\end{subequations}
The vector $\boldsymbol{J}$ represents the vector of the reaction fluxes, where $J_\rho$ is the net rate at which reaction $\rho$ occurs in the forward direction; similarly, $\boldsymbol{I}^Y$ is the exchange flux vector, with $I_y^Y$ specifying the net rate at which species $y$ is transferred from the chemostat to the CRN.
For example, in the GG network (Fig.~\ref{fig:1}a), the concentrations of the internal species G6P and the external species ATP evolve according to
\begin{subequations}
    \label{eq:crn_dynamics}
    \begin{equation}
        d_t[\text{G6P}] = J_1^\text{GG} + J_2^\text{GG} -J_3^\text{GG},
    \end{equation}
    \begin{equation}
        d_t[\text{ATP}] = -J_1^\text{GG} - J_4^\text{GG}  + J_9^\text{GG} + J_{12}^\text{GG} +   J_{14}^\text{GG} + {I}_\text{ATP}^Y.
    \end{equation}
\end{subequations}
Under the stationary conditions considered in this paper, the flux satisfies:
\begin{subequations}
    \begin{equation}
        d_t{\boldsymbol{x}} = 0 \implies \mathbb{S}^X \boldsymbol{J} = 0,
    \label{eq:stationary conditions X}
    \end{equation}
    \begin{equation}
        d_t{\boldsymbol{y}}= 0 \implies \mathbb{S}^Y\boldsymbol{J} = -\boldsymbol{I}^Y.
            \label{eq:stationary conditions Y}
    \end{equation}
\end{subequations}
We note that fixing either the external concentrations to $\boldsymbol{y}$ or the exchange fluxes to  $\boldsymbol{I}^Y$ leads to equivalent stationary conditions.

We emphasize that the results presented in this paper also apply to compartmentalized CRNs, provided each compartment is well-stirred. In such cases, the same chemical species in different compartments are treated as distinct.
\subsection{CRN Cycles}
\label{Sect:Cycles}
Given an open CRN, a cycle $\boldsymbol{\psi}$ is defined as any sequence of reactions that upon completion does not change the concentrations of the internal species $X$.
For example, the sequences of reactions +1 and -2 in the GG network (Fig.~\ref{fig:1}a) and +3, +5, and -2 in the RF network (Fig.~\ref{fig:1}c) form cycles whose sole effect is ATP hydrolysis. 
Mathematically, $\boldsymbol{\psi}$ is a vector in the space of reactions whose entries, $\psi_\rho$, denote the number of times each reaction $\rho$ occurs. This vector satisfies the condition 
\begin{equation}
    \mathbb{S}^X\,\boldsymbol{\psi} = 0,
\end{equation}
which ensures that the concentrations of species $X$ remain unchanged. The previous two cycles in mathematical form read 
\begin{equation}
\boldsymbol{\psi}^\text{GG} =\,
\begin{array}{l}
\scriptstyle \textcolor{gray}{1 }\\
\scriptstyle \textcolor{gray}{2}
\end{array}
\begin{pmatrix}
+1 \\
-1 
\end{pmatrix},
\quad \quad 
\boldsymbol{\psi}^\text{RF} =\, 
\begin{array}{l}
\scriptstyle \textcolor{gray}{2}\\
\scriptstyle \textcolor{gray}{3} \\
\scriptstyle \textcolor{gray}{5}
\end{array}
\begin{pmatrix}
-1 \\
+1 \\
+1 
\end{pmatrix},
\end{equation}
where only the nonzero entries are displayed for simplicity. The matrix $\mathbb{S}^Y$ quantifies how cycle $\boldsymbol{\psi}$ affects the $Y$ species. Accordingly, we classify cycles as \emph{external} if they alter $Y$-species concentrations, $\mathbb{S}^Y \boldsymbol{\psi}\ne0$,
and as \emph{internal} if they do not,  $\mathbb{S}^Y \boldsymbol{\psi}=0$. 
Cycles $\boldsymbol{\psi}^\text{GG} $ and $\boldsymbol{\psi}^\text{RF} $ are both external since they hydrolyze ATP, whereas
\begin{equation}
\boldsymbol{\psi'}^\text{GG} =\, 
\begin{array}{l}
\scriptstyle \textcolor{gray}{1}\\
\scriptstyle \textcolor{gray}{2}\\
\scriptstyle \textcolor{gray}{4} \\
\scriptstyle \textcolor{gray}{5}
\end{array}
\begin{pmatrix}
+1\\
-1 \\
-1 \\
+1 
\end{pmatrix}
\end{equation}
is an example of an internal cycle. 
The full space of cycles can be described using a basis that consists of two distinct sets of vectors,  $\{\boldsymbol{\phi}_i\}\cup\{\boldsymbol{\phi}_\epsilon\}$. Here, $\{\boldsymbol{\phi}_i\}$ is a basis for the internal cycles and $\{\boldsymbol{\phi}_\epsilon\}$ is a set of independent external cycles, referred to as \emph{emergent} cycles~\cite{polettiniIrreversibleThermodynamicsOpen2014a}. The number of $\{\boldsymbol{\phi}_i\}$ is given by $|i|=$dim(ker($\mathbb{S}$)), while the number of emergent cycles is given by $|\epsilon|=$dim(ker($\mathbb{S}^X$)) $-$ dim(ker($\mathbb{S}$)). The selection of these two sets is in general arbitrary. For example, the GG network in Fig.~\ref{fig:1}a has one internal cycle, 
\begin{equation}
\boldsymbol{\phi}_{i_1}^\text{GG} =\, 
\begin{array}{l}
\scriptstyle \textcolor{gray}{1}\\
\scriptstyle \textcolor{gray}{2}\\
\scriptstyle \textcolor{gray}{4} \\
\scriptstyle \textcolor{gray}{5}
\end{array}
\begin{pmatrix}
+1\\
-1 \\
-1 \\
+1 
\end{pmatrix},
\end{equation}
and three emergent cycles that can be chosen as follows:
\begin{equation}
\boldsymbol{\phi}_{\epsilon_1}^\text{GG} = 
\, 
\begin{array}{l}
\scriptstyle \textcolor{gray}{1}\\
\scriptstyle \textcolor{gray}{2}
\end{array}
\begin{pmatrix}
+1\\
-1 \\
\end{pmatrix}, \,
\boldsymbol{\phi}_{\epsilon_2}^\text{GG}  =
\begin{array}{l}
\scriptstyle \textcolor{gray}{12}\\
\scriptstyle \textcolor{gray}{13}\\
\scriptstyle \textcolor{gray}{14}
\end{array}
\begin{pmatrix}
+1\\
-1 \\
-1
\end{pmatrix},\,
\boldsymbol{\phi}_{\epsilon_3}^\text{GG}  =
\begin{array}{l}
\scriptstyle \textcolor{gray}{1}\\
\scriptstyle \textcolor{gray}{3}\\
\scriptstyle \textcolor{gray}{4}\\
\scriptstyle \textcolor{gray}{6}\\
\scriptstyle \textcolor{gray}{7}\\
\scriptstyle \textcolor{gray}{8}\\
\scriptstyle \textcolor{gray}{9}\\
\scriptstyle \textcolor{gray}{10}\\
\scriptstyle \textcolor{gray}{11}\\
\scriptstyle \textcolor{gray}{12}
\end{array}
\begin{pmatrix}
+1\\
+1 \\
+1 \\
+1\\
+1 \\
+2\\
+2\\
+2 \\
+2\\
+2
\end{pmatrix}
.
\label{eq:Emerg. cycles GG}
\end{equation}
The RF network in Fig.~\ref{fig:1}c has only three emergent cycles and a possible choice for them is
\begin{equation}
\boldsymbol{\phi}_{\epsilon_1}^\text{RF} = 
\, 
\begin{array}{l}
\scriptstyle \textcolor{gray}{1}\\
\scriptstyle \textcolor{gray}{3}\\
\scriptstyle \textcolor{gray}{6}
\end{array}
\begin{pmatrix}
+1\\
+1 \\
+1
\end{pmatrix}, \,
\boldsymbol{\phi}_{\epsilon_2}^\text{RF}  =
\begin{array}{l}
\scriptstyle \textcolor{gray}{1}\\
\scriptstyle \textcolor{gray}{2}\\
\scriptstyle \textcolor{gray}{4}
\end{array}
\begin{pmatrix}
+1\\
+1 \\
+1
\end{pmatrix},\,
\boldsymbol{\phi}_{\epsilon_3}^\text{RF}  =
\begin{array}{l}
\scriptstyle \textcolor{gray}{4}\\
\scriptstyle \textcolor{gray}{5}\\
\scriptstyle \textcolor{gray}{6}
\end{array}
\begin{pmatrix}
+1\\
+1 \\
-1 
\end{pmatrix}
.
\label{eq:Emerg. cycles RF}
\end{equation}
Finally, we note that any stationary flux $\boldsymbol{J}$ is also a cycle since it leaves the concentrations of the $X$ species unchanged, satisfying Eq.~\eqref{eq:stationary conditions X}. Therefore, it can be decomposed using the basis $\{\boldsymbol{\phi}_i\}\cup\{\boldsymbol{\phi}_\epsilon\}$: 
\begin{equation}
        \boldsymbol{J} = \sum_i J_i\boldsymbol{\phi}_i + \sum_\epsilon J_\epsilon\boldsymbol{\phi}_\epsilon.
        \label{eq:J decomposition_cycles}
\end{equation}

\subsection{Atomic Composition Matrix}
\label{Sect:Atomic composition matrix}

The atomic composition matrix $\mathbb{A}$ encodes the atomic-level structure of chemical species---a level of detail generally unavailable from the stoichiometric matrix alone.  This matrix is essential for defining chemical processes, as discussed later in Sect.~\ref{Sect:Chemical processes}. For example, each element $\mathbb{A}_{a,z}$ denotes the number of atoms of type $a$ in the chemical species $z$. To incorporate molecular charge, $\mathbb{A}$ includes an additional row, where each entry represents the charge of the corresponding species.  
The concentration of atoms $a$ in the system is given by
\begin{equation}
    [a]  = \sum_{x}\mathbb{A}_{a,x} \,[x] + \sum_{y}\mathbb{A}_{a,y}\, [y]. 
\end{equation}
Since the net change in chemical species due to any reaction $\rho$, quantified by $\mathbb{S}_\rho$, conserves atomic counts and charge, it follows that
\begin{equation}
    \mathbb{A}_a \, \mathbb{S}_\rho = 0,
    \label{eq: rho preserves mass charge}
\end{equation}
where $\mathbb{A}_a$ is the row corresponding to atom $a$. Given that this holds for every reaction, we get:
\begin{equation}
    \mathbb{A}_a\, \mathbb{S} = 0.
\end{equation}
 Thus, the row vectors of $\mathbb{A}$ lie in the cokernel (left null space) of $\mathbb{S}$.
However, $\mathbb{A}$ may not fully span the cokernel, and its rows can also be linearly dependent~\cite{Muller_What_makesCRN_chemical_2022}. The latter happens, for example, when two atomic types always appear in the same ratio across all species. 
Analogously to the stoichiometric matrix, the atomic composition matrix $\mathbb{A}$ can be partitioned into internal and external species blocks:
\begin{equation}
    \mathbb{A} = \left(\mathbb{A}^X \,\,\, \mathbb{A}^Y \right),
\end{equation}
 Since emergent cycles $\boldsymbol{\phi}_e$ generate effective reactions involving only $Y$ species, given by $\mathbb{S}^Y \boldsymbol{\phi}_e$, they satisfy:
\begin{equation}
    \mathbb{A}^Y\mathbb{S}^Y\boldsymbol{\phi}_e = 0.
\end{equation}
In what follows, we focus exclusively on effective reactions among external species $Y$ since, under stationary conditions, they are the only ones being produced or consumed. 
Accordingly, only $\mathbb{A}^Y$, which encodes the atomic composition of external species, is relevant.
Therefore, a coarse-graining of internal species or reactions is admissible, provided the resulting network gives rise to effective reactions among external species that preserve mass and charge---as in the RF network in Fig.~\ref{fig:1}c. In addition, the coarse-graining must maintain thermodynamic consistency, as discussed in the next section.

\subsection{Local Validity of the Second Law}
\label{Sect:Local validity of the Second law}
In the following, we assume that the CRN is embedded in an isothermal and isobaric reservoir (e.g. the water solution in biochemistry) and that the second law is valid for every reaction $\rho$, that is:
\begin{equation}
    -J_\rho\Delta_\rho G \ge 0.
    \label{eq:Local second law}
\end{equation}
$\Delta_\rho G = \boldsymbol{\mu}\cdot \mathbb{S}_\rho$ is the Gibbs free energy change associated with reaction $\rho$, with $\boldsymbol{\mu} = (\dots,\mu_i,\dots)$ the vector of chemical potentials. This assumption is well known for elementary reactions, but it also encompasses any non-elementary reaction resulting from a coarse-graining of multiple elementary reactions into a single \emph{emergent} cycle~\cite{avanziniCircuitTheoryChemical2023}. This includes, for example, any enzymatic reactions occurring in the cytosol~\cite{wachtelThermodynamicallyConsistentCoarse2018} as well as the effective reactions in the RF network (Fig.~\ref{fig:1}c) corresponding to glycolysis (reaction 1) and TCA-cycle + Oxidative phosphorylation (reaction 4)~\cite{wachtelFreeenergyTransductionChemical2022}.

\section{Chemical Processes}
\label{Sect:Chemical processes}
Under stationary conditions, the CRN draws in multiple $Y$ species from the surroundings, transforms them through chemical reactions, and releases newly formed $Y$ species back into the environment. At its core, this operation redistributes atoms among external species while exchanging heat with the thermal reservoir. The chemical processes introduced here, particularly the elementary ones, provide a systematic framework for characterizing this redistribution.


\subsection{Definition}
A chemical process is defined as an effective reaction among the $Y$ species that is stoichiometrically balanced with respect to mass and charge. 
Since these quantities are conserved by all reactions in the CRN, Eq.~\eqref{eq: rho preserves mass charge}, each external cycle $\boldsymbol{\psi}$ identifies a valid process $\boldsymbol{p} = \mathbb{S}^Y\boldsymbol{\psi}$. 
For example, the emergent cycle $\boldsymbol{\phi}_{\epsilon_1}^\text{GG}$ in Eq.~\eqref{eq:Emerg. cycles GG} realizes
ATP hydrolysis, 
    \begin{equation}
         \boldsymbol{p}_{\epsilon_1}^\text{GG} :\, \text{ATP} + \text{H}_2\text{O} \rightarrow \text{ADP} + \text{P}_i + \text{H}^+,
        \label{eq:p_e1^GG}
        \end{equation}
and the emergent cycle $\boldsymbol{\phi}_{\epsilon_1}^\text{RF}$ from Eq.~\eqref{eq:Emerg. cycles RF} realizes the process
        \begin{equation}
           \boldsymbol{p}_{\epsilon_1}^\text{RF}:\,\,
            \begin{array}{c}
            \text{Glc} + 2\text{ADP} + 2\text{P}_i + 2\text{H}^+ \\
            \downarrow \\
            2\text{Eth} + 2\text{CO}_2 + 2\text{ATP} + 2\text{H}_2\text{O}
            \end{array}\, ,
            \label{eq:p_e1^RF}
        \end{equation}
which corresponds to the fermentation of glucose into ethanol that yields 2 ATP molecules.
Formally, a process $\boldsymbol{p}$ is a vector in the space of $Y$ species, where each entry $p_y$ denotes the number of $y$ molecules consumed or produced by that process. We can rewrite mathematically the two processes above as
\begin{equation}
     \boldsymbol{p}_{\epsilon_1}^\text{GG}   =
\begin{array}{l}
\scriptstyle \textcolor{gray}{\text{ATP}}\\
\scriptstyle \textcolor{gray}{\text{H}_2\text{O}}\\
\scriptstyle \textcolor{gray}{\text{ADP}}\\
\scriptstyle \textcolor{gray}{\text{P}_i}\\
\scriptstyle \textcolor{gray}{\text{H}^+}\\
\end{array}
\begin{pmatrix}
-1\\
-1 \\
+1 \\
+1 \\
+1 
\end{pmatrix},\quad
 \boldsymbol{p}_{\epsilon_1}^\text{RF} = 
\begin{array}{l}
\scriptstyle \textcolor{gray}{\text{Glc}}\\
\scriptstyle \textcolor{gray}{\text{Eth}}\\
\scriptstyle \textcolor{gray}{\text{CO}_2}\\
\scriptstyle \textcolor{gray}{\text{ATP}}\\
\scriptstyle \textcolor{gray}{\text{H}_2\text{O}}\\
\scriptstyle \textcolor{gray}{\text{ADP}}\\
\scriptstyle \textcolor{gray}{\text{P}_i}\\
\scriptstyle \textcolor{gray}{\text{H}^+}\\
\end{array}
\begin{pmatrix}
-1\\
+2 \\
+2 \\
+2 \\
+2 \\
-2 \\
-2 \\
-2 
\end{pmatrix},
\label{eq:processes ex vector form}
\end{equation}
where we display only the nonzero entries for conciseness. 
The fact that any chemical process conserves atomic counts and charge translates into the condition: 
\begin{equation}
    \mathbb{A}^Y\, \boldsymbol{p} = 0,
\end{equation}
where $\mathbb{A}^Y$ is the atomic composition matrix reduced to the external species from Sect.~\ref{Sect:Atomic composition matrix}.

\emph{Elementary} processes constitute a special class of processes representing the most basic interconversions among the $Y$ species. These are defined as processes that cannot be further decomposed into subprocesses (see Appendix~\ref{app:Elementary processes} for more details including how to compute them). For example, $\boldsymbol{p}_{\epsilon_1}^\text{GG}$ is elementary, while $\boldsymbol{p}_{\epsilon_1}^\text{RF}$ is not. The reason is that the latter carries out two subprocesses: the conversion of glucose to ethanol, $\text{Glc} \rightarrow 2\text{Eth} + 2\text{CO}_2$, and ATP synthesis, $ 2\text{ADP} + 2\text{P}_i + 2\text{H}^+ \rightarrow 2\text{ATP} + 2\text{H}_2\text{O}$, both of which are elementary.

We emphasize that the definition of processes only depends on the atomic composition of the $Y$ species. The processes that a CRN can actually execute (under stationary conditions) form a subset of all possible processes:
\begin{equation}
    \Pi_\text{CRN} = \{ \mathbb{S}^Y\boldsymbol{J}\},
    \label{eq:Processes executable by CRN}
\end{equation}
where $\boldsymbol{J}$ is any stationary flux. In Appendix~\ref{app:Characterization of PI_CRN}, we characterize $\Pi_\text{CRN}$, demonstrating that the processes implemented by the emergent cycles form a basis for this set.

\subsection{Complete and Independent Sets of Processes}
\label{Sect:Complete and independent sets of processes}

This section aims to identify a set of processes that provides a \emph{univocal} description of matter flow through the CRN, that is, of the intake and release of $Y$ species quantified by the exchange flux vector $\boldsymbol{I}^Y$ in Eq.~\eqref{eq:stationary conditions Y}. 
Establishing such a set will be essential for the later discussion of transduction in Sect.~\ref{Sect: Requirements for a Set of Processes for transduction}. To serve this purpose, the set of processes must meet two fundamental requirements: completeness and linear independence.
We call a set $\{\boldsymbol{p}\}$ \emph{complete} if linear combinations of its elements can represent any process realizable by the CRN (\emph{i.e.}, any process in $\Pi_\text{CRN}$ defined in Eq.~\eqref{eq:Processes executable by CRN}). Such completeness enables the decomposition of   $\boldsymbol{I}^Y$:
\begin{equation}
    -\boldsymbol{I}^Y = \mathbb{S}^Y \boldsymbol{J} = \sum_{\{\boldsymbol{p}\}} \mathcal{I}_{\boldsymbol{p}}\,\boldsymbol{p},\\
    \label{eq:I decomposition_processes}
\end{equation}
where $\mathcal{I}_{\boldsymbol{p}}$ denotes the steady state rate at which process $\boldsymbol{p}$ occurs:  $\mathcal{I}_{\boldsymbol{p}}>0$ when the process proceeds forward  and $\mathcal{I}_{\boldsymbol{p}}<0$ when it proceeds backward. If the set $\{\boldsymbol{p}\}$ is both complete and linearly independent, this decomposition becomes unique. Substituting Eq.~\eqref{eq:I decomposition_processes} into the entropy production expression for the chemostats then yields a corresponding unique decomposition of the entropy production:
\begin{equation}
    \begin{split}
          \dot \Sigma = \boldsymbol{\mu}_Y\cdot \boldsymbol{I}^Y =   -\sum_{\{\boldsymbol{p}\}} \mathcal{I}_{\boldsymbol{p}}  \,\Delta_{\boldsymbol{p}}  G.     
    \end{split}
    \label{eq:Sigma decomposition_processes}
\end{equation}
Here, $\boldsymbol{\mu}_Y$ represents the vector of chemical potentials of the $Y$ species, which specifies the operating conditions, and  $\Delta_{\boldsymbol{p}} G = \boldsymbol{\mu}_Y\cdot\boldsymbol{p}$ is the Gibbs free energy change caused by process $\boldsymbol{p}$. This Gibbs free energy change is equivalent to the sum of the products' chemical potentials minus the sum of the reactants' chemical potentials. For any given CRN, it is always possible to identify a complete and independent set of processes. Examples for the GG and RF networks will be presented later in Eqs.~\eqref{eq:Complete-independent set of processes GG} and \eqref{eq:Complete-independent set of processes RF}, together with the corresponding decompositions of $ -\boldsymbol{I}^Y$ and $\dot \Sigma$ in Eqs.~\eqref{eq: I decomposition_processes GG and RF} and \eqref{eq: Sigma decomposition_processes GG and RF}.

\section{Transduction Efficiency}
\label{Sect:Transduction Efficiency}

In Sect.~\ref{Sect:Chemical processes}, we introduced the notion of chemical processes as a means to decompose the flux of $Y$ species through the CRN, $\boldsymbol{I}^Y$, into mass- and charge-preserving components (see Eq.~\eqref{eq:I decomposition_processes}). Given that energy transduction is, at its core, the transfer of free energy among chemical processes---processes going thermodynamically downhill drive others uphill---the first step to analyze transduction is to identify a suitable set of processes, $\mathcal{T}$, that we use to interpret and quantify these free energy transfers. In the simplest case where there is a single input and a single output, the choice of $\mathcal{T}$ is straightforward. However, when multiple processes are involved, several choices for $\mathcal{T}$ become possible.

To reduce this ambiguity, in Sect.~\ref{Sect: Requirements for a Set of Processes for transduction}, we first establish the essential criteria that any admissible set $\mathcal{T}$ must satisfy and examine the consequences of an improper choice. 
Yet, even after these constraints are imposed, multiple valid sets $\mathcal{T}$ typically remain---each potentially yielding a different definition of transduction efficiency. Therefore, in Sect.~\ref{Sect:Procedure for Selecting the Set of Processes}, we introduce a physically motivated procedure for unambiguously selecting $\mathcal{T}$. This procedure relies on defining a reference equilibrium environment. In Sect.~\ref{Sect:Reference Equilibrium}, we explain how to define it in the context of CRNs and show how a reference equilibrium naturally identifies a unique set of processes. Additionally, we outline how this procedure is closely connected to the thermodynamic concept of exergy, whose precise meaning and relevance will be detailed in that same section.

Once $\mathcal{T}$ has been established, we turn to the definition of transduction efficiency in Sect.~\ref{Sect:Def of efficiency}.
There, we explain how to distinguish input and output processes, formally express efficiency in Eq.~\eqref{eq:Definition efficiency}, and discuss its inherently relative nature.

\subsection{Suitable Set of Processes $\mathcal{T}$ for Transduction}
\label{Sect: Requirements for a Set of Processes for transduction}
A set of processes $\mathcal{T}$ suitable for analyzing transduction must satisfy three key requirements: It must be $i)$ complete (see Sect.~\ref{Sect:Complete and independent sets of processes}), $ii)$ linearly independent, and $iii)$ composed exclusively of elementary processes.

Completeness ensures that $\mathcal{T}$ can describe any possible CRN's net effect on the chemostatted species, while linear independence  ensures that transduction efficiency is unambiguously defined. We now illustrate why. Consider the following set of processes for the RF network that satisfies neither of these two requirements:  
\begin{equation}
    \{  \boldsymbol{p}_{\text{Glc}_R},\,\boldsymbol{p}_{\text{Eth}_R},\boldsymbol{p}_{\text{Glc}_F}\},
\end{equation}
with 
\begin{subequations}
    \label{eq:incomplete set of processes}
    \begin{equation}
        \boldsymbol{p}_{\text{Glc}_R} :\,  \text{Glc} + 6\text{O}_2 \rightarrow 6\text{CO}_2 + 6\text{H}_2\text{O},
    \end{equation}
    \begin{equation}
        \boldsymbol{p}_{\text{Eth}_R} :\, \text{Eth} + 3\text{O}_2 \rightarrow 2\text{CO}_2 + 3\text{H}_2\text{O},
    \end{equation}
       \begin{equation}
        \boldsymbol{p}_{\text{Glc}_F} :\,\text{Glc} \rightarrow 2\text{Eth} + 2\text{CO}_2.
    \end{equation}
\end{subequations} 
$\boldsymbol{p}_{\text{Glc}_R},\,\boldsymbol{p}_{\text{Eth}_R}$ represent glucose and ethanol respiration, while  $\boldsymbol{p}_{\text{Glc}_F}$ glucose fermentation.
This set is problematic for two reasons: it cannot describe ATP consumption or production within the CRN, and it exhibits linear dependence, as 
$\boldsymbol{p}_{\text{Glc}_R} =  \boldsymbol{p}_{\text{Glc}_F} +2 \boldsymbol{p}_{\text{Eth}_R}$. 
Such dependence leads to multiple possible interpretations of the same process run by the CRN:
For example, glucose respiration might be equivalently described as glucose fermentation followed by ethanol respiration (performed twice). These different interpretations yield distinct decompositions of $-\boldsymbol{I}^Y$ and entropy production, Eqs.~\eqref{eq:I decomposition_processes} and \eqref{eq:Sigma decomposition_processes}, and different definitions of transduction efficiency, as will become clear later through Eq.~\eqref{eq:Definition efficiency}.

While the theory applies generally, requirement $iii)$ is motivated by the following physical observation: Elementary processes constitute the smallest fundamental units by which free energy can be generated or dissipated in chemostats. 
Moreover, transduction occurring at the level of elementary processes may become undetectable when analyzed through non-elementary ones. As an example of this, consider the simple CRN:
 \begin{equation}
  \ce{ $A + B$ <=>[+1][-1]  $AB$ <=>[+2][-2] $CD$ <=>[+3][-3] $C+ D$}.
  \label{eq: TC CRN example}
\end{equation}
where the $Y$ species are $A,\, B,\, C,$ and $D$. This CRN has a single emergent cycle, corresponding to the sequence of reactions +1, +2, and +3, which realizes the process 
\begin{equation}
    \boldsymbol{p} :  A + B \rightarrow  C + D.
    \label{eq: TC CRN example global process}
\end{equation}
Suppose that knowledge of the mass and charge of the $Y$ species reveals that $\boldsymbol{p}$ consists of two elementary processes:
\begin{equation}
    \boldsymbol{p} = \boldsymbol{p}' + \boldsymbol{p}'',
\end{equation}
with
\begin{subequations}
\label{eq: TC CRN example elementary processes}
    \begin{equation}
    \boldsymbol{p}' :  A \rightarrow  C  \qquad \Delta G' = \mu_C -\mu_A <0,
\end{equation}
\begin{equation}
    \boldsymbol{p}'' :  B \rightarrow  D \qquad  \Delta G'' = \mu_D -\mu_B >0.
\end{equation}
\end{subequations}
Assuming a stationary flux from left to right, the CRN clearly performs transduction when analyzed in terms of $ \boldsymbol{p}'$ and $ \boldsymbol{p}''$: A portion of the free energy consumed by $\boldsymbol{p}'$ is utilized to drive the thermodynamically unfavorable process $\boldsymbol{p}''$ with an efficiency $\eta = |\Delta G''|/|\Delta G'|$.
However, when the same CRN is described in terms of the single, non-elementary process $\boldsymbol{p}$, which satisfies requirements $i)$ and $ii)$, this transduction becomes invisible since $\boldsymbol{p}$ is the only process at play.

\subsection{Reference Equilibrium: Associated Processes and Chemical Exergies.}
\label{Sect:Reference Equilibrium}

In this subsection, we first explain how a reference equilibrium environment is established for open CRNs. We then show that species out of equilibrium with respect to this reference naturally determine a set of processes that satisfy the first two criteria from Sect.~\ref{Sect: Requirements for a Set of Processes for transduction}: completeness and linear independence. Finally, we clarify how the Gibbs free energies associated with these processes are connected to the thermodynamic concept of exergy. These results will serve as the conceptual basis for the procedure developed in the next section, aimed at constructing the set $\mathcal{T}$ used to analyze transduction. 

Setting a reference equilibrium for the $Y$ species interacting through the CRN amounts to specifying their equilibrium chemical potentials, denoted by the vector $\boldsymbol{\mu}_{Y}^\text{eq}$. To be in equilibrium, these must satisfy the condition that, for every process $\boldsymbol{p}$ executable by the CRN, $\boldsymbol{p} \in \Pi_\text{CRN}$, the associated Gibbs free energy change vanishes: 
\begin{equation}
    \Delta_{\boldsymbol{p}} G^\text{eq} = \boldsymbol{\mu}_Y^\text{eq} \cdot \boldsymbol{p} = 0.
    \label{eq: Equilibrium condition Delta G}
\end{equation}
Since the number of independent processes in $\Pi_\text{CRN}$ corresponds to the number of independent emergent cycles, $|\epsilon|$ (see Appendix~\ref{app:Characterization of PI_CRN}), this imposes $|\epsilon|$ independent constraints on $\boldsymbol{\mu}_Y^\text{eq}$. Consequently, only $|Y_P| = |Y|  - |\epsilon|$ equilibrium chemical potentials need to be specified to fully determine $\boldsymbol{\mu}_Y^\text{eq}$. The corresponding subset $Y_P \subset Y$ is referred to as the set of \emph{potential} species since it defines the equilibrium chemical potentials of all $Y$.
In contrast, the remaining species $Y_F \subset Y$, with $|Y_F| = |\epsilon|$, are referred to as the \emph{force} species. These species generally possess nonequilibrium chemical potentials relative to $\boldsymbol{\mu}_Y^\text{eq}$, thereby driving the CRN away from equilibrium. Since both the GG and RF networks feature three independent emergent cycles, Eq.~\eqref{eq:Emerg. cycles GG} and ~\eqref{eq:Emerg. cycles RF}, three force species are required in each case.
A possible choice used in the subsequent transduction analysis is:
\begin{equation}
    Y_F^\text{GG} = \{\text{Glc},\text{GTP},\text{ATP} \},\, \quad Y_F^\text{RF} =  \{\text{Glc},\text{Eth},\text{ATP}\}.
\end{equation}
Once the classification of $Y$ species is achieved, each force species $y_F\in Y_F$ identifies a nonequilibrium process $\boldsymbol{p}_F$ that exclusively involves itself and $Y_P$, \emph{i.e.}, mathematically: $p_{F,y} \neq 0$ only when $y = y_F$ or $y\in Y_P$ (see Appendix~\ref{App:Processes associated with Y_F}). Since each $\boldsymbol{p}_F$ converts a different $y_F$ into $Y_P$, these processes are all independent. In addition, in Appendix~\ref{App:Processes associated with Y_F},  we show that the set $\mathcal{P}=\{\boldsymbol{p}_F\}$ is also complete. For the GG network, the set $Y_F^\text{GG}$  
identifies:
\begin{equation}
    \begin{array}{c}
           Y_F^\text{GG} = \{\text{Glc},\text{GTP},\text{ATP} \} \\
            \Downarrow \\
           \mathcal{P}^\text{GG} = \{\boldsymbol{p}_{\text{Glc}\to\text{Pyr}},\, \boldsymbol{p}_\text{GTP},\, \boldsymbol{p}_\text{ATP} \},
    \end{array}
    \label{eq:Complete-independent set of processes GG}
\end{equation}
which correspond to the conversion of glucose into pyruvate, GTP and ATP hydrolysis. 
Similarly, for the RF network, the set $Y_F^\text{RF}$  
identifies: 
\begin{equation}
    \begin{array}{c}
           Y_F^\text{RF} =  \{\text{Glc},\text{Eth},\text{ATP}\} \\
            \Downarrow \\
           \mathcal{P}^\text{RF} = \{\boldsymbol{p}_{\text{Glc}_R},\boldsymbol{p}_{\text{Eth}_R},\,\boldsymbol{p}_\text{ATP}\},
    \end{array}
    \label{eq:Complete-independent set of processes RF}
\end{equation}
which correspond to glucose and ethanol respiration, and ATP hydrolysis. 
Explicitly, these processes are (Figs.~\ref{fig:1}b and \ref{fig:1}d):
\begin{subequations}
    \label{eq:All GG RF processes}
    \begin{equation}
    \boldsymbol{p}_{\text{Glc}\to\text{Pyr}} :\, \text{Glc} + 2\text{NAD}^+ \rightarrow 2\text{Pyr} + 2\text{NADH} + 2\text{H}^+,
    \end{equation}
    \begin{equation}
        \boldsymbol{p}_\text{GTP} :\, \text{GTP} + \text{HCO}_3 \rightarrow \text{GDP} + \text{P}_i + \text{CO}_2,
    \end{equation}
        \begin{equation}
        \boldsymbol{p}_\text{ATP} :\, \text{ATP} + \text{H}_2\text{O} \rightarrow \text{ADP} + \text{P}_i + \text{H}^+,
    \end{equation}
    \begin{equation}
        \boldsymbol{p}_{\text{Glc}_R} :\,  \text{Glc} + 6\text{O}_2 \rightarrow 6\text{CO}_2 + 6\text{H}_2\text{O},
    \end{equation}
    \begin{equation}
        \boldsymbol{p}_{\text{Eth}_R} :\, \text{Eth} + 3\text{O}_2 \rightarrow 2\text{CO}_2 + 3\text{H}_2\text{O}.
    \end{equation}
\end{subequations} 
Since both $  \mathcal{P}^\text{GG} $ and $  \mathcal{P}^\text{RF} $ are complete and independent, they can be used to uniquely decompose the flux of $Y$ species transferred from the CRN to the chemostats, Eq.~\eqref{eq:I decomposition_processes}: 
\begin{subequations}
\label{eq: I decomposition_processes GG and RF}
    \begin{equation}
    -\boldsymbol{I}^{Y,\text{GG}} = \mathcal{I}_{\text{Glc}\to\text{Pyr}} \,\boldsymbol{p}_{\text{Glc}\to\text{Pyr}} +  \mathcal{I}_\text{GTP}\, \boldsymbol{p}_{\text{GTP}} +  \mathcal{I}_\text{ATP}\, \boldsymbol{p}_{\text{ATP}},
\end{equation}
    \begin{equation}
    -\boldsymbol{I}^{Y,\text{RF}} = \mathcal{I}_{\text{Glc}_R} \,\boldsymbol{p}_{\text{Glc}_R} +  \mathcal{I}_{\text{Eth}_R}\, \boldsymbol{p}_{\text{Eth}_R} +  \mathcal{I}_\text{ATP}\, \boldsymbol{p}_{\text{ATP}},
\end{equation}
\end{subequations}
The Gibbs free energy changes associated with these processes, $\Delta_{F} G = \boldsymbol{\mu}_Y \cdot \boldsymbol{p}_F$, quantify the independent nonequilibrium thermodynamic forces acting on the CRN. In particular, when $\Delta_{F} G = 0$, the chemical potential of species $y_F$ matches its equilibrium value, as determined by the chemical potentials of the potential species $Y_P$. Similarly, $ \mathcal{P}^\text{GG} $ and $ \mathcal{P}^\text{RF} $ can be used to uniquely decompose the entropy production as in Eq.~\eqref{eq:Sigma decomposition_processes}:
\begin{subequations}
\label{eq: Sigma decomposition_processes GG and RF}
    \begin{equation}
    \begin{split}
          \dot \Sigma^{\text{GG}} = & - \mathcal{I}_{\text{Glc}\to\text{Pyr}}  \,\Delta_{\text{Glc}\to\text{Pyr}}  G - \mathcal{I}_{\text{GTP}}  \,\Delta_{{\text{GTP}}}  G\\
          &- \mathcal{I}_{\text{ATP}}  \,\Delta_{{\text{ATP}}}  G,
     \end{split}
    \label{eq:GG Sigma decomposition_processes}
\end{equation}
    \begin{equation}
    \dot \Sigma^{\text{RF}} = -\mathcal{I}_{\text{Glc}_R} \,\Delta_{\text{Glc}_R} G-  \mathcal{I}_{\text{Eth}_R}\, \Delta_{\text{Eth}_R} G -  \mathcal{I}_\text{ATP}\, \Delta_{\text{ATP}} G,
    \label{eq:RF Sigma decomposition_processes}
\end{equation}
\end{subequations}
where
\begin{subequations}
    \label{eq:Delta_G processes explicit GG RF}
    \begin{equation}
        \Delta_{\text{Glc}\to\text{Pyr}} G = 2\mu_{\text{Pyr}} +2  \mu_{\text{NADH}} +  2\mu_{\text{H}^+} -  \mu_{\text{Glc}} - 2\mu_{\text{NAD}^+}, 
    \end{equation}
        \begin{equation}
        \Delta_{\text{GTP} } G = \mu_{\text{GDP}} +  \mu_{\text{P}_i} +  \mu_{\text{CO}_2} -  \mu_{\text{GTP}} - \mu_{\text{HCO}_3},
    \end{equation}
    \begin{equation}
        \Delta_{\text{ATP} } G = \mu_{\text{ADP}} +  \mu_{\text{P}_i} +  \mu_{\text{H}^+} -  \mu_{\text{ATP}} - \mu_{\text{H}_2\text{O}},
    \end{equation}
        \begin{equation}
        \Delta_{\text{Glc}_R } G = 6\mu_{\text{CO}_2} + 6 \mu_{\text{H}_2\text{O}} -  \mu_{\text{Glc}} - 6 \mu_{\text{O}_2},
    \end{equation}
    \begin{equation}
        \Delta_{\text{Eth}_R } G = 2\mu_{\text{CO}_2} + 3 \mu_{\text{H}_2\text{O}} -  \mu_{\text{Eth}} - 3 \mu_{\text{O}_2}. 
    \end{equation}
\end{subequations}
Throughout the following analysis, we assume that all of the above Gibbs free energy changes are negative, which is reasonable under typical physiological conditions (see Table~\ref{Tab:Standard physiological conditions}).

\textit{Exergy Interpretation.}
We first define the concept of exergy for a steady-state system coupled to multiple reservoirs not in mutual equilibrium~\cite{Bejan2016AdvancedThermodynamics,szargut_exergy_Analysis_1988}.  The exergy of a reservoir is defined as the maximum useful work obtainable through the transfer of conserved quantities (energy, matter, or charge) between that reservoir and the subset of reservoirs defining the reference equilibrium environment. For example, in the chemical context, once the reference equilibrium is fixed via the choice of potential species $Y_P$, the exergy of each force species $y_F \in Y_F$ becomes well-defined. It corresponds to the Gibbs free energy change of the
associated process $\boldsymbol{p}_F$: 
\begin{equation}
    |\Delta_{F} G| = |\boldsymbol{\mu}_Y \cdot \boldsymbol{p}_F|.
\end{equation}
This value quantifies the maximum useful work extractable from the nonequilibrium concentration of $y_F$. In particular, if $\boldsymbol{p}_F$ is rescaled so that it consumes exactly one molecule of $y_F$ (i.e., $p_{F, y_F} = 1$), then $|\Delta_{F} G|$ corresponds to the chemical exergy per molecule of $y_F$.

In the next section, by analyzing transduction in terms of the processes $\{\boldsymbol{p}_F\}$, we adhere to the physically meaningful definition of efficiency as the ratio between output and input exergy. This approach generalizes to chemistry the method used to define efficiency for thermal engines connected to multiple heat baths~\cite{Bejan2016AdvancedThermodynamics}. In the simpler case of thermal engines, a single ``potential" temperature suffices to establish the reference equilibrium, as energy is the only exchanged quantity. In contrast,  chemical transduction involves the exchange of multiple species with the surroundings, requiring multiple potential species $Y_P$ to define the reference equilibrium.

Finally, because exergy values depend on the chosen reference equilibrium, the resulting transduction efficiency may also inherit the same contextual dependence. This arbitrariness will be commented on in the next section~\ref{Sect:Procedure for Selecting the Set of Processes}.

\subsection{Procedure for Selecting the Set of Processes $\mathcal{T}$ for Transduction.}
\label{Sect:Procedure for Selecting the Set of Processes}
Here, building on Sect.~\ref{Sect:Reference Equilibrium}, we introduce the procedure for selecting the set of processes, $\mathcal{T}$, that meets the three requirements outlined in Sect.~\ref{Sect: Requirements for a Set of Processes for transduction}. We begin with a schematic overview of the procedure before delving into a discussion of its individual steps.
\begin{enumerate}
    \item Define a reference equilibrium and determine the corresponding set of processes $\{\boldsymbol{p}_F\}$ linked to the force species, as explained in Sect.~\ref{Sect:Reference Equilibrium}. This set satisfies the first two requirements: completeness and independence.
    \item Next, verify whether all processes in $\{\boldsymbol{p}_F\}$ are elementary (see Appendix~\ref{app:Elementary processes}):
    \begin{enumerate}
        \item  If all processes are elementary: we simply set $\mathcal{T} = \{\boldsymbol{p}_F\}$.
        \item  If not: Consider a non-elementary process $\boldsymbol{p}_F$ and extract from it an elementary subprocess $\boldsymbol{p}$ (see Appendix~\ref{App:Existence of an elementary subprocess}). Then, redefine the CRN to include an additional `virtual' reaction that realizes $\boldsymbol{p}$ and repeat the procedure from step 1. This updated CRN will feature an additional emergent cycle, necessitating the selection of an additional force species when defining a reference equilibrium.
    \end{enumerate}
\end{enumerate}
If we apply step 1 to the GG and RF networks choosing $Y_F^\text{GG} = \{\text{Glc},\text{GTP},\text{ATP} \}$ and $Y_F^\text{RF} =  \{\text{Glc},\text{Eth},\text{ATP}\}$, one obtains the two sets of processes in Eqs.~\eqref{eq:Complete-independent set of processes GG} and \eqref{eq:Complete-independent set of processes RF}. It can be verified that these processes are all elementary (see Appendix~\ref{app:Elementary processes}). Therefore, the two sets for transduction are:
\begin{subequations}
\label{eq: set of processes transduction GG and RF}
    \begin{equation}
     \mathcal{T}^\text{GG} = \{\boldsymbol{p}_{\text{Glc}\to\text{Pyr}},\, \boldsymbol{p}_\text{GTP},\, \boldsymbol{p}_\text{ATP} \},
     \label{eq: set of processes transduction GG}
\end{equation}
    \begin{equation}
    \mathcal{T}^\text{RF} = \{\boldsymbol{p}_{\text{Glc}_R},\boldsymbol{p}_{\text{Eth}_R},\,\boldsymbol{p}_\text{ATP}\}.
    \label{eq: set of processes transduction RF}
\end{equation}
\end{subequations}
Regarding the arbitrariness of step 1 in selecting the reference equilibrium, the specific case study can offer guidance. A practical general criterion is to designate the more abundant species in the environment as $Y_P$. The rationale behind this choice is that, due to their abundance, the concentrations of these species---and, consequently, the reference equilibrium they define---remain effectively unchanged, even over the long time scales (not considered here) where chemostats equilibrate. Additionally, for interpretability, it is often useful to classify as $Y_F$ those species that are naturally regarded as the system's `fuel species'. 
We also observe that many different choices for the reference equilibrium result in the same set of processes, thus resolving part of the arbitrariness. For instance, in the GG network, the following choices for the force species: $\{\text{Glc},\text{GTP},\text{ATP} \}$, $\{\text{NAD}^+,\text{GTP},\text{ATP} \}$, $\{\text{Glc},\text{HCO}_3,\text{ATP} \}$, and $\{\text{Glc},\text{GTP},\text{H}_2\text{O} \}$, all lead to the same $\mathcal{T}^\text{GG}$ in Eq.~\eqref{eq: set of processes transduction GG}.

Step 2 ensures that transduction is fully resolved by analyzing it in terms of elementary processes. In particular, it applies to CRNs that cannot execute certain elementary processes, \emph{i.e.}, these processes do not belong to $\Pi_\text{CRN}$, Eq.~\eqref{eq:Processes executable by CRN}. This situation arises, for example, in tightly coupled CRNs, such as the one described in Eq.~\eqref{eq: TC CRN example}, where elementary processes ($\boldsymbol{p}'$ and $\boldsymbol{p}''$) always occur together and in a fixed ratio. In the case of tightly coupled CRNs, step 1 yields a single non-elementary $\boldsymbol{p}_F$, which in our case corresponds to $ \boldsymbol{p}$ in Eq.~\eqref{eq: TC CRN example global process}. 
To proceed further, step 2 (b) is applied: an elementary subprocess of $\boldsymbol{p}$---for instance, $\boldsymbol{p}'$---is extracted and introduced as a virtual reaction in the CRN. This results in the following updated CRN:
 \begin{equation}
 \begin{split}
      & \ce{ $A + B$ <=>[+1][-1]  $AB$ <=>[+2][-2] $CD$ <=>[+3][-3] $C+ D$},\\
  & \qquad \qquad \quad \ce{ $A $ <=>[+4][-4]  $C$}.
 \end{split}
  \label{eq: TC CRN example + virtual reaction}
\end{equation}
Then, repeating step 1 on this modified CRN and choosing, for example, $Y_F =\{A,B\}$ produces the correct set of elementary processes $\mathcal{T} = \{\boldsymbol{p}',\,\boldsymbol{p}''\}$. The intuitive idea behind step 2 (b) is that,
by adding a virtual reaction corresponding to an elementary process, it allows the CRN to resolve non-elementary processes into their constituent elementary ones.\\

\subsection{Definition of Efficiency}
\label{Sect:Def of efficiency}

Once a proper set of processes $\mathcal{T}$ has been identified, we can define the transduction efficiency $\eta$. Here, the key conceptual step is determining which processes qualify as input and which as output. Notably, this classification, $\mathcal{T}= \mathcal{T}_\text{in} \cup \mathcal{T}_\text{out} $, is not solely dictated by the network topology and operating conditions but also by the specific CRN stationary state. In particular, the reaction fluxes within the network determine whether a process proceeds thermodynamically downhill (input) or uphill (output). This classification naturally emerges from entropy production, where positive contributions correspond to input processes and negative contributions to output processes. 
We first illustrate the procedure for the two CRN examples before providing the general definition in Eq.~\eqref{eq:Definition efficiency}.\\

\paragraph{Examples.}
From Eq.~\eqref{eq:GG Sigma decomposition_processes}, the stationary entropy production of the GG network decomposed in terms of $\mathcal{T}^\text{GG}$ reads
\begin{equation}
\begin{split}
             \dot \Sigma^{\text{GG}} = & - \mathcal{I}_{\text{Glc}\to\text{Pyr}}  \,\Delta_{\text{Glc}\to\text{Pyr}}  G - \mathcal{I}_{\text{GTP}}  \,\Delta_{{\text{GTP}}}  G\\
          &- \mathcal{I}_{\text{ATP}}  \,\Delta_{{\text{ATP}}}  G.
\end{split}
          \label{eq:Reported GG Sigma decomposition_processes}
\end{equation}
Given our assumption $\Delta_{F} G<0$ for all processes, the sign of the contributions in Eq.~\eqref{eq:Reported GG Sigma decomposition_processes} is determined by the direction of the chemostat fluxes,
$\mathcal{I}_{\text{Glc}\to\text{Pyr}},\, \mathcal{I}_{\text{GTP}},$ and $ \mathcal{I}_{\text{ATP}}$, which depends on the chemical potentials $\boldsymbol{\mu}_Y$ and the reactions' kinetics.
We can consider various scenarios. In the case of gluconeogenesis, ATP and GTP are consumed to generate glucose, meaning that $\mathcal{I}_{\text{Glc}\to\text{Pyr}} < 0,\,\mathcal{I}_{\text{GTP}} > 0,\, \mathcal{I}_{\text{ATP}} >0$. If we denote the input--output split using the notation
\begin{equation}
    s: \mathcal{T}_\text{in}\to \mathcal{T}_\text{out},
    \label{eq:s notation}
\end{equation} 
then for gluconeogenesis one has:
\begin{subequations}
    \label{eq:Gluconeogenesis}
    \begin{equation}
            s^\text{Gluc}: \,\{ \boldsymbol{p}_\text{GTP},\, \boldsymbol{p}_\text{ATP}\} \to \{ -\boldsymbol{p}_{\text{Glc}\to\text{Pyr}}\},
            \label{eq:s Gluconeogenesis}
    \end{equation}
     \begin{equation}
        \eta(s^\text{Gluc}) = \frac{ \mathcal{I}_{\text{Glc}\to\text{Pyr}}\Delta_{\text{Glc}\to\text{Pyr}}  G   }{ 
     -  \mathcal{I}_\text{GTP} \Delta_\text{GTP} G  -  \mathcal{I}_\text{ATP} \Delta_\text{ATP} G },
     \label{eq:eta Gluconeogenesis}
    \end{equation}
\end{subequations}
where the transduction efficiency is defined as the ratio of the output to the input terms. 
In the case of glycolysis, glucose is consumed to generate ATP. Depending on whether GTP is degraded or produced, two possible scenarios arise: 
\begin{subequations}
     \label{eq:Glycolisis 1}
    \begin{equation}
        s^\text{Glyc}_1: \,\{\boldsymbol{p}_{\text{Glc}\to\text{Pyr}},\,\boldsymbol{p}_\text{GTP}\} \to \{ - \boldsymbol{p}_\text{ATP}\}, 
         \label{eq:s Glycolisis 1}
    \end{equation}
     \begin{equation}
        \eta(s^\text{Glyc}_1) = \frac{ \mathcal{I}_\text{ATP} \Delta_\text{ATP} G}{- \mathcal{I}_{\text{Glc}\to\text{Pyr}}\Delta_{\text{Glc}\to\text{Pyr}}  G -  \mathcal{I}_\text{GTP} \Delta_\text{GTP} G }
         \label{eq:eta Glycolisis 1}
    \end{equation}
\end{subequations}
and
\begin{subequations}
     \label{eq:Glycolisis 2}
    \begin{equation}
        {s}^\text{Glyc}_2: \,\{\boldsymbol{p}_{\text{Glc}\to\text{Pyr}} \} \to \{ -\boldsymbol{p}_\text{ATP}, -\boldsymbol{p}_\text{GTP}\}, 
         \label{eq:s Glycolisis 2}
    \end{equation}
     \begin{equation}
        {\eta}(s^\text{Glyc}_2)= \frac{  \mathcal{I}_\text{ATP} \Delta_\text{ATP} G+ \mathcal{I}_\text{GTP} \Delta_\text{GTP} G }{- \mathcal{I}_{\text{Glc}\to\text{Pyr}} \Delta_{\text{Glc}\to\text{Pyr}} G }.
         \label{eq:eta Glycolisis 2}
    \end{equation}
\end{subequations}
Similarly, from Eq.~\eqref{eq:RF Sigma decomposition_processes}, the entropy production of the RF network can be decomposed as
\begin{equation}
          \dot \Sigma^{\text{RF}} = -\mathcal{I}_{\text{Glc}_R} \,\Delta_{\text{Glc}_R} G-  \mathcal{I}_{\text{Eth}_R}\, \Delta_{\text{Eth}_R} G -  \mathcal{I}_\text{ATP}\, \Delta_{\text{ATP}} G. 
           \label{eq:Reported RF Sigma decomposition_processes}
\end{equation}
 If the CRN uses glucose as the carbon source, oxydizing it to either ethanol or CO$_2$, one has $\mathcal{I}_{\text{Glc}_R}>0$, $\mathcal{I}_{\text{Eth}_R}<0 $, and $\mathcal{I}_{\text{ATP}}<0 $, leading to: 
\begin{subequations}
    \label{eq:Glucose RF}
    \begin{equation}
        s^\text{Glc}:\{
\boldsymbol{p}_{\text{Glc}_R} \}\to\{-\boldsymbol{p}_\text{ATP},-\boldsymbol{p}_{\text{Eth}_R}\},
     \label{eq:s Glucose RF}
    \end{equation}
    \begin{equation}
        \eta(s^\text{Glc}) = \frac{ \mathcal{I}_{\text{Eth}_R}\Delta_{\text{Eth}_R}  G   +  \mathcal{I}_\text{ATP} \Delta_\text{ATP} G }{ -  \mathcal{I}_{\text{Glc}_R}\Delta_{\text{Glc}_R}G }.
        \label{eq:eta Glucose RF}
    \end{equation}
\end{subequations}
Using this last example as an illustration,  we highlight an important feature of the framework: the flexibility to selectively include only a portion of the output processes when defining efficiency. This is particularly valuable when certain outputs, though thermodynamically significant, are not utilized by the system---their stored free energy is ultimately dissipated through external reactions, making it effectively irrecoverable.
For example, consider a scenario in which a yeast (RF network) ferments glucose into ethanol, and bacteria in the surrounding environment consume that ethanol as part of subsequent stages in the ecological chain. From the yeast's perspective, the free energy stored in ethanol is effectively lost, as the yeast does not reuse it. As a result, the efficiency with which the yeast captures the glucose's free energy relies solely on the ATP produced during fermentation. This efficiency is therefore smaller and equal to
\begin{align}
     {\eta^*}(s^\text{Glc}) &= \frac{ \mathcal{I}_\text{ATP} \Delta_\text{ATP} G }{ -  \mathcal{I}_{\text{Glc}_R}\Delta_{\text{Glc}_R} G } \label{eq:eta Glucose RF useful output}\\ 
     & \le \frac{ \mathcal{I}_{\text{Eth}_R}\Delta_{\text{Eth}_R}  G   +  \mathcal{I}_\text{ATP} \Delta_\text{ATP} G }{ -  \mathcal{I}_{\text{Glc}_R}\Delta_{\text{Glc}_R} G} = \eta(s^\text{Glc}).\nonumber     
\end{align}
To formalize this approach, depending on the specific system, one can partition the full set of output processes into two disjoint subsets: \emph{target} output processes (or `useful' processes)---those whose stored free energy can be harnessed by the system at a later stage---and \emph{non-recoverable} output processes, whose stored free energy is effectively lost. This distinction allows us to define the target efficiency, denoted by $\eta^*$, which quantifies the system's effectiveness at converting input free energy solely into its designated target outputs. We formalize it later in Eq.~\eqref{eq:Definition efficiency useful output}.
We note that while focusing on a subset of output processes ensures that the resulting target efficiency remains below 1, the same cannot be said for input processes. Input processes represent free energy costs in the system, and neglecting any of these costs could result in an unphysical efficiency bigger than 1.\\

\paragraph{General Procedure}
Having presented specific examples, we now outline the general prescription for defining transduction efficiency, followed by a discussion highlighting its inherently relative nature.
\begin{enumerate}
    \item Select the set of processes $\mathcal{T}$ (see Sect.~\ref{Sect:Procedure for Selecting the Set of Processes}) and use it to decompose the entropy production:
    \begin{equation}
    \begin{split}
        \dot\Sigma = - \sum_{\boldsymbol{p}\in \mathcal{T}} \, \mathcal{I}_{\boldsymbol{p}} \Delta_{\boldsymbol{p}} G >0.
         \label{eq:Sigma decomposition_processes second law}
    \end{split}
    \end{equation}
    (We assume nonzero fluxes and therefore nonzero dissipation). 
    \item Split this set into input and output processes, $\mathcal{T} = \mathcal{T}_\text{in}\cup \mathcal{T}_\text{out}$, depending on the sign of the individual contributions:
     \begin{equation}
    \begin{split}
        \dot\Sigma &=  \underbrace{\sum_{\boldsymbol{p}'\in \mathcal{T}_\text{in}} -\, \mathcal{I}_{\boldsymbol{p}'} \Delta_{\boldsymbol{p}'} G}_{\dot\Sigma_\text{in}(s)> 0}\, + \underbrace{\sum_{\boldsymbol{p}''\in \mathcal{T}_\text{out}} -\,\mathcal{I}_{\boldsymbol{p}''} \Delta_{\boldsymbol{p}''} G}_{ \dot\Sigma_\text{out}(s)\le 0} >0. 
          \label{eq:Sigma decomposition_processes split}
    \end{split}
    \end{equation}
     $\dot\Sigma_\text{in}(s)$ is the sum of positive contributions, which represent input fluxes of free energy, while $\dot\Sigma_\text{out}(s)$ is the sum of negative contributions, which represent output fluxes of free energy. Again, $s:\mathcal{T}_\text{in}\to \mathcal{T}_\text{out} $ denotes the input-output split.
    \item If $\mathcal{T}_\text{out} = \emptyset $, no transduction is occurring. Otherwise, the transduction efficiency is defined as
    \begin{equation}
     \eta(s) = \frac{-\dot\Sigma_\text{out}(s) }{\dot\Sigma_\text{in}(s) } < 1.
        \label{eq:Definition efficiency}
    \end{equation}
    \item Depending on the system under analysis, restrict the set of output processes only to the useful ones, $\mathcal{T}_\text{out}^*\subset\mathcal{T}_\text{out}$: 
    \begin{equation}
     \eta^*(s) = \frac{-\dot\Sigma_\text{out}^*(s) }{\dot\Sigma_\text{in}(s) } \le \eta(s) < 1,
        \label{eq:Definition efficiency useful output}
    \end{equation} 
    where $\dot\Sigma_\text{out}^*(s) = - \sum_{\boldsymbol{p}''\in \mathcal{T}_\text{out}^*} \,\mathcal{I}_{\boldsymbol{p}''} \Delta_{\boldsymbol{p}''} G$.
\end{enumerate}
This procedure makes it clear that  $\eta$ is not an absolute quantity. Its definition depends on the set of processes $\mathcal{T}$ chosen for analyzing transduction, which can vary depending on the selected reference environment, on the sign of $-\mathcal{I}_{\boldsymbol{p}}\Delta_{\boldsymbol{p}}G$, which determines whether a process is classified as input or output, and finally on which output processes are considered `useful'.
Moreover, the transduction efficiency associated with a CRN can change when the same CRN is analyzed as part of a larger network. 

The first reason for this is that a larger network may capture additional dissipation associated with transduction. For example, in the GG network, all reactions are catalyzed by specific enzymes that, in reality, have a finite lifespan and undergo periodic degradation and re-synthesis. This cycle of degradation and re-synthesis results in additional consumption of ATP and GTP. While this cost of maintaining enzymes is not included in the GG network we are considering, it could be captured by a larger CRN.

A second reason is that expanding the network can introduce new external species, potentially requiring a redefinition of the reference equilibrium. In particular, species previously classified as potential species may be reclassified as force species, and vice versa. Such a redefinition can, in turn, alter the set of processes used to analyze transduction, $\mathcal{T}\to \mathcal{T}'$, thereby changing how efficiency is defined.
For instance, consider the GG network as part of a larger CRN that includes the full respiratory metabolism.
In this expanded network, new external species such as $\text{O}_2$ and $\text{CO}_2$ (together with H$_2$O) would be included in $Y_P'$. Instead, pyruvate would turn into a force species---that is, a species from which free-energy can still be extracted.
As a result, while transduction occurring in the GG network was previously analyzed in terms of the conversion of glucose into pyruvate, $\boldsymbol{p}_{\text{Glc}\to\text{Pyr}}\in \mathcal{T}$, it would now be analyzed in terms of glucose and pyruvate respiration, $\boldsymbol{p}_{\text{Glc}_R},\, \boldsymbol{p}_{\text{Pyr}_R}\in \mathcal{T}'$, where:
\begin{equation}
    \boldsymbol{p}_{\text{Pyr}_R} :\, \text{Pyr} + 3\text{O}_2+  \text{NADH} + \text{H}^+ \rightarrow 3\text{CO}_2 + 3\text{H}_2\text{O} +\text{NAD}^+.
\end{equation}
We stress that this relativity in the notion of efficiency is intrinsic and unavoidable. Recognizing it allows for a more informed analysis of transduction.

\section{Transduction Gears}
\label{Sect:Gears}
\begin{figure*}
    \centering
    \includegraphics[width=1.0\linewidth]{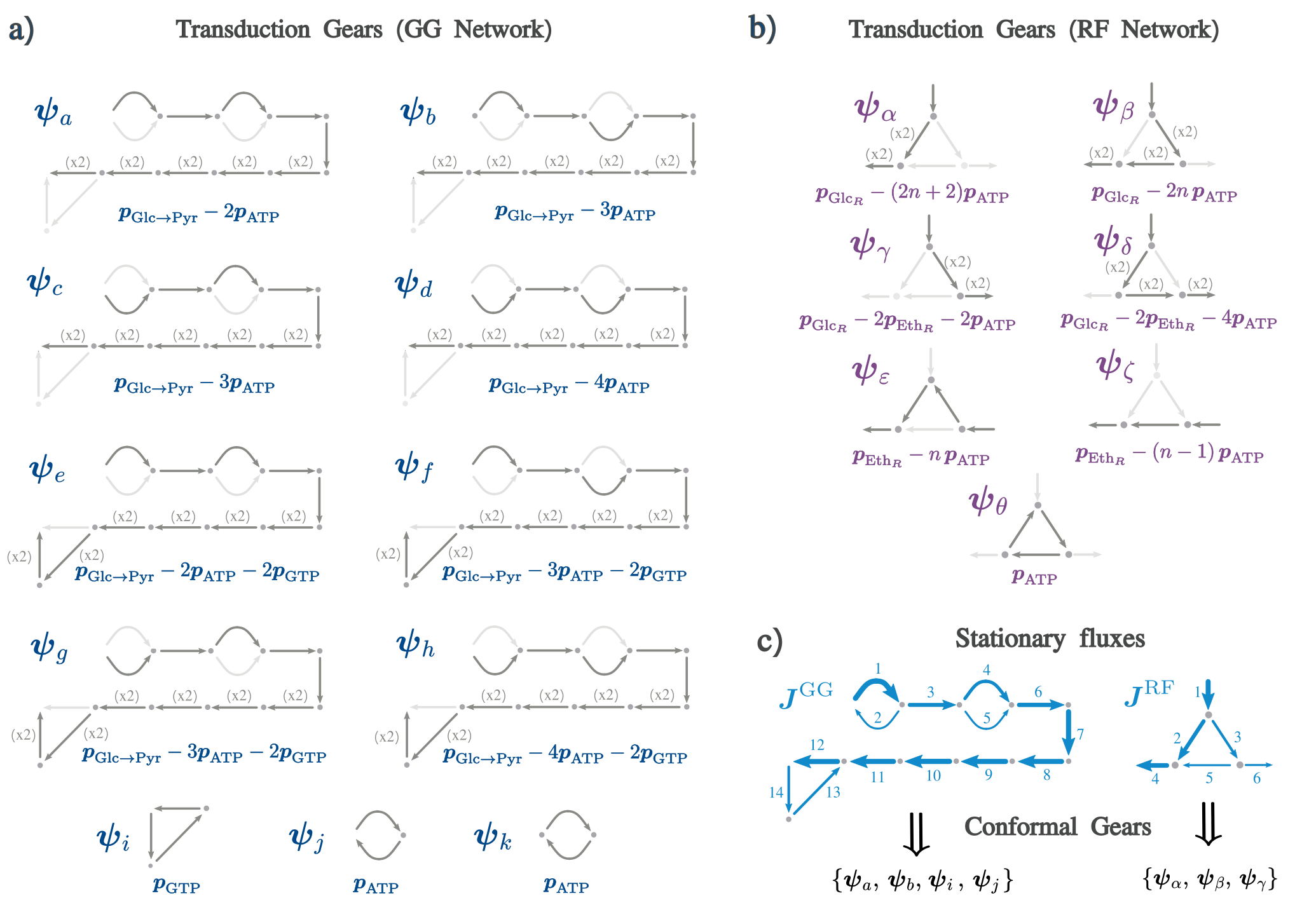}
    \caption{\textbf{CRN gears for transduction.} \textbf{a) + b)}  Schematic representations of transduction gears for the CRNs shown in Fig.~\ref{fig:1}. These gears correspond to the external EFMs and intuitively represent all possible ``reaction paths" that pass through the internal species, leaving them unaltered. Due to the nonlinear nature of the CRNs, certain reactions occur twice.
     Below each gear, its net effect on the chemostatted species is shown, decomposed in terms of the processes in Fig.~\ref{fig:1}b and \ref{fig:1}d. Gears $\boldsymbol{\psi}_i$, $\boldsymbol{\psi}_j$, and $\boldsymbol{\psi}_k$ of the GG network and gear $\boldsymbol{\psi}_\theta$ of the RF network are non-transducing since they exclusively hydrolyze either ATP or GTP. The remaining eight gears of the GG network convert glucose to pyruvate synthesizing a variable number of ATP and GTP molecules. In the RF network, $\boldsymbol{\psi}_\alpha$ and $\boldsymbol{\psi}_\beta$ perform glucose respiration; $\boldsymbol{\psi}_\gamma$ and $\boldsymbol{\psi}_\delta$ perform glucose fermentation; and $\boldsymbol{\psi}_\varepsilon$ and $\boldsymbol{\psi}_\zeta$ perform ethanol respiration.  \textbf{c)}  Examples of stationary fluxes (with arrow thickness proportional to magnitude) in the GG and RF networks and the respective subset of conformal (active) gears, \emph{i.e.}, the gears that are aligned with the flux direction for every reaction.}
    \label{fig:2}
\end{figure*}
In~\cite{bilancioni2025gears}, we formalized the notion of chemical gears as distinct transduction pathways within CRNs, highlighting their crucial role in optimizing a CRN’s operation from an efficiency standpoint. However, our previous analysis was limited to the case of a single input and a single output process. 
In Sect.~\ref{Sect:Def gear efficiency}, we develop the concept of gears for multi-process transduction. In this broader setting, we show that a single gear can be associated with multiple efficiencies depending on the input-output classification, characterize the gears that can contribute to transduction, and examine the implications of reversing the direction of transduction. Sect.~\ref{Sect:Conformal gears} identifies, for a given flux configuration within the CRN, which gears are `active'. Since the notion of gears is derived from \emph{elementary flux modes} (EFMs)~\cite{zanghellini_EFM_nutshell_2013}, we begin by briefly recalling their definition.


\subsection{Elementary Flux Modes (EFMs)}
\label{Sect:EFMs}

An elementary flux mode (EFM) is a special kind of cycle (see Sect.~\ref{Sect:Cycles}) that uses a minimal \emph{set} of reactions. The reactions $\rho$ used by a cycle $\boldsymbol{\psi}$ are the reactions for which $\psi_\rho \ne 0$, and a set of reactions is considered minimal if removing any one of them makes it impossible to form a valid cycle with the remaining ones. We underscore that
EFMs are a purely topological concept and are defined up to a multiplicative factor. Intuitively, in linear CRNs, EFMs correspond to the reaction paths of the associated graph that pass through the internal species while leaving them unaltered. In nonlinear CRNs, they extend this concept to hypergraphs. The GG and the RF networks have 11 and 7 EFMs, respectively, displayed in Fig.~\ref{fig:2}a and \ref{fig:2}b.
EFMs can be identified using specialized algorithms~\cite{Terzer_efmtool_2008}. However, as network size increases, the number of EFMs grows exponentially, making their complete enumeration computationally challenging for large metabolic networks~\cite{zanghellini_EFM_nutshell_2013}.
Like cycles, EFMs can be classified as \emph{internal} if they do not alter the $Y$ species and as \emph{external} in case they do.

\subsection{Gears and Gear's Efficiency}
\label{Sect:Def gear efficiency}
Transduction gears of a CRN are defined as its external EFMs~\cite{bilancioni2025gears} and, as such, they are a topological property of the network. CRNs with a single external EFM are referred to as single-gear CRNs, while those with multiple external EFMs are termed multi-gear CRNs.
Within this framework, both the GG and RF networks are examples of multi-gear CRNs: In particular, all their EFMs, shown in Fig.~\ref{fig:2}a and b, are external and thus qualify as gears.
We emphasize that gears, like EFMs, are defined up to a multiplicative factor.
To each gear, we can assign an efficiency, defined as the free energy it generates in the chemostats via the output processes divided by the free energy it consumes from the chemostats via the input processes. We start with an example and then
provide the general definition in Eq.~\eqref{eq:Gear efficiency definition}. 
Consider gear $\boldsymbol{\psi}_h$ of the GG network in Fig.~\ref{fig:2}a. This gear realizes a process, $\boldsymbol{p}_h = \mathbb{S}^Y \boldsymbol{\psi}_h$, that can be decomposed in terms of the set $\mathcal{T}^\text{GG}$, Eq.~\eqref{eq: set of processes transduction GG}:
\begin{equation}
    \boldsymbol{p}_h = \boldsymbol{p}_{\text{Glc}\to\text{Pyr}} - 2\,\boldsymbol{p}_\text{GTP} - 4\,\boldsymbol{p}_\text{ATP}.
\end{equation}
In words, $\boldsymbol{p}_h$ converts glucose to pyruvate, synthesizing at the same time 2 GTP and 4 ATP molecules.
Analogously, the Gibbs free energy change caused by gear $\boldsymbol{\psi}_h$ can be decomposed as
\begin{equation}
    \Delta_{h} G = \boldsymbol{\mu}_Y \cdot   \boldsymbol{p}_h =\Delta_{\text{Glc}\to\text{Pyr}} G - 2\, \Delta_\text{GTP} G -4\, \Delta_\text{ATP} G.
\end{equation}
The efficiency $\eta_h$ of this gear depends on the input-output split. For example, for gluconeogenesis,  $s^\text{Gluc}$ in Eq.~\eqref{eq:s Gluconeogenesis}, $-\boldsymbol{\psi}_h$ consumes a free energy $-2\, \Delta_\text{GTP} G -4\, \Delta_\text{ATP} G $ from the chemostats via the input processes $\boldsymbol{p}_\text{GTP}$ and $\boldsymbol{p}_\text{ATP}$ and generates a free energy $-\Delta_{\text{Glc}\to\text{Pyr}} G$ in the chemostats via the output process $-\boldsymbol{p}_{\text{Glc}\to\text{Pyr}}$. Therefore, its efficiency is
\begin{equation}
    \eta_h(s^\text{Gluc}) =  \frac{-\Delta_{\text{Glc}\to\text{Pyr}} G }{-2\, \Delta_\text{GTP} G -4\, \Delta_\text{ATP} G}.
\end{equation}
Similarly, for the input-output splits that correspond to glycolysis, $s^\text{Glyc}_1$ and $s^\text{Gluc}_2$ in Eqs.~\eqref{eq:s Glycolisis 1} and \eqref{eq:s Glycolisis 2}, one has 
\begin{subequations}
    \begin{equation}
        \eta_h(s^\text{Glyc}_1)  =  \frac{ -4\, \Delta_\text{ATP} G}{-\Delta_{\text{Glc}\to\text{Pyr}} G + 2\, \Delta_\text{GTP} G },
    \end{equation}
    \begin{equation}
        \eta_h(s^\text{Glyc}_2)  =  \frac{ -2\, \Delta_\text{GTP} G  - 4\, \Delta_\text{ATP} G}{-\Delta_{\text{Glc}\to\text{Pyr}} G}.
    \end{equation}
\end{subequations}

    Having shown an example, we now present the general procedure for defining the efficiency of a gear.
\begin{enumerate}
    \item If $\boldsymbol{\psi}_{\mathfrak{g}}$ is the gear under consideration, decompose the associated process, $\boldsymbol{p}_{\mathfrak{g}} = \mathbb{S}^Y \boldsymbol{\psi}_{\mathfrak{g}}$, in terms of $\mathcal{T}$, the set of processes selected for analyzing transduction (see Sect.~\ref{Sect:Procedure for Selecting the Set of Processes}):
    \begin{equation}
        \boldsymbol{p}_{\mathfrak{g}} = \sum_{\boldsymbol{p}\in \mathcal{T}} m^{\mathfrak{g}}_{\boldsymbol{p}} \,\boldsymbol{p}.
        \label{eq:Gear process decomposition_processes}
    \end{equation}
    $m^{\mathfrak{g}}_{\boldsymbol{p}}$ is the number of times that $\boldsymbol{\psi}_{\mathfrak{g}}$ implements the elementary process ${\boldsymbol{p}}$ in the forward (backward if $m^{\mathfrak{g}}_{\boldsymbol{p}}<0$) direction and depends solely on the CRN's topology. In Fig.~\ref{fig:2}a and \ref{fig:2}b, for example, we report this decomposition for each gear of the GG and RF networks.
    \item Use the above decomposition to rewrite in the same way the Gibbs free energy change, $\Delta_{\mathfrak{g}} G$,  associated with gear $\boldsymbol{\psi}_{\mathfrak{g}}$:
    \begin{equation}
     \Delta_{\mathfrak{g}} G = \boldsymbol{\mu}_Y \cdot   \boldsymbol{p}_{\mathfrak{g}} = \sum_{\boldsymbol{p}\in \mathcal{T}} m^{\mathfrak{g}}_{\boldsymbol{p}} \, \Delta_{\boldsymbol{p}} G.
     \label{eq:Gear delta_G decomposition_processes}
    \end{equation}
    Then, divide the terms according to the particular input-output split considered, $s: \mathcal{T}_\text{in}\to \mathcal{T}_\text{out}$.
    \begin{equation}
     \Delta_{\mathfrak{g}} G = \underbrace{ \sum_{\boldsymbol{p}'\in \mathcal{T}_\text{in}} m^{\mathfrak{g}}_{\boldsymbol{p}'} \Delta_{\boldsymbol{p}'} G}_{\Delta_{\mathfrak{g}} G_\text{in}(s)}\, +  \underbrace{\sum_{\boldsymbol{p}''\in \mathcal{T}_\text{out}}m^{\mathfrak{g}}_{\boldsymbol{p}''} \Delta_{\boldsymbol{p}''} G}_{\Delta_{\mathfrak{g}} G_\text{out}(s)} >0.
     \label{eq:Gear delta G decomposition_processes split}
    \end{equation}
    $\Delta_{\mathfrak{g}} G_\text{in}(s)$ (resp. $\Delta_{\mathfrak{g}} G_\text{out}(s)$) represents the overall free energy change caused by gear $\boldsymbol{\psi}_{\mathfrak{g}}$ in the chemostats via the input (resp. output) processes. 
    \item Define the efficiency of gear $\boldsymbol{\psi}_{\mathfrak{g}}$ as
    \begin{equation}
        \eta_{\mathfrak{g}}(s) = \frac{-\Delta_{\mathfrak{g}} G_\text{out}(s)}{\Delta_{\mathfrak{g}} G_\text{in}(s)}.
        \label{eq:Gear efficiency definition}
    \end{equation}
    Importantly, this efficiency is independent of whether one considers $\boldsymbol{\psi}_{\mathfrak{g}}$ or $-\boldsymbol{\psi}_{\mathfrak{g}}$.
    \item  Depending on the system under analysis, restrict the set of output processes to consider only the `useful' ones, $\mathcal{T}_\text{out}^*\subset\mathcal{T}_\text{out}$, as discussed in Sect.~\ref{Sect:Def of efficiency}:
    \begin{equation}
        \eta_{\mathfrak{g}}^*(s) = \frac{-\Delta_{\mathfrak{g}} G_\text{out}^*(s)}{\Delta_{\mathfrak{g}} G_\text{in}(s)},
         \label{eq:Gear efficiency definition useful output}
    \end{equation}
    where $\Delta_{\mathfrak{g}} G_\text{out}^*(s) = \sum_{\boldsymbol{p}''\in \mathcal{T}_\text{out}^*}m^{\mathfrak{g}}_{\boldsymbol{p}''} \Delta_{\boldsymbol{p}''} G$.
\end{enumerate}
Based on the value of $\Delta_{\mathfrak{g}} G_\text{in}(s)$ and $\Delta_{\mathfrak{g}} G_\text{out}(s)$, which are function of  the $Y$ chemical potentials, $\eta_{\mathfrak{g}}$ in Eq.~\eqref{eq:Gear efficiency definition} can be negative, bigger than one, or even infinite. \emph{Transducing} gears are gears for which $0<\eta_{\mathfrak{g}}(s)<1$: They can transfer free energy from input to output with a thermodynamically feasible efficiency. 
Whether a gear is transducing depends on the operating conditions, given by $\boldsymbol{\mu}_Y$, and the input-output split $s$. 
Gears that engage only with a single process in $\mathcal{T}$ are never transducing. This is the case, for example, of gears $\boldsymbol{\psi}_i$, $\boldsymbol{\psi}_j$, $\boldsymbol{\psi}_k$ in the GG network (Fig.~\ref{fig:2}a) and gear $\boldsymbol{\psi}_\theta$ in the RF network (Fig.~\ref{fig:2}b): They only realize either ATP or GTP hydrolysis, which makes their efficiencies either zero or infinite depending on whether this single process counts as input (futile gear) or as output. In Appendix~\ref{app:Upper bound}, we show that every non-transducing gear, when present, always has a detrimental effect on the transduction efficiency. 

Finally, we observe that reversing the direction of transduction, where the input becomes the output and vice versa, 
\begin{equation}
    \begin{array}{c}
            s: \mathcal{T}_\text{in} \to \mathcal{T}_\text{out} \\
            \Downarrow \\
            s^{-1}: \mathcal{T}_\text{out} \to \mathcal{T}_\text{in}
    \end{array}\, ,
\end{equation}
results in an inversion of gear efficiencies:
\begin{equation}
    \eta_{\mathfrak{g}}(s^{-1})  = \frac{1}{\eta_{\mathfrak{g}}(s)}.
    \label{eq:Efficiency reverse transduction}
\end{equation}
This inversion flips the roles of gears: the heaviest ones become the lightest, while the lightest become the heaviest. For example, $s^\text{Gluc}$, in Eq.~\eqref{eq:s Gluconeogenesis}, and
${s}^\text{Glyc}_2$, in Eq.~\eqref{eq:s Glycolisis 2}, are the reversed input-output splits of each other. If we consider the efficiencies of gears $\boldsymbol{\psi}_a$ and $ \boldsymbol{\psi}_d$ (Fig.~\ref{fig:2}a), they are given in the two cases by
\begin{equation}
   \eta_a(s^\text{Gluc}) = \frac{-\Delta_{\text{Glc}\to\text{Pyr}}  G   }{ -2\, \Delta_\text{ATP} G }\,\,> \,\,  \eta_d(s^\text{Gluc}) = \frac{-\Delta_{\text{Glc}\to\text{Pyr}}  G   }{ -4\, \Delta_\text{ATP} G }
\end{equation}
and 
\begin{equation}
     \eta_a(s^\text{Glyc}_2) = \frac{- 2\, \Delta_\text{ATP} G }{-\Delta_{\text{Glc}\to\text{Pyr}}  G} \,\,< \,\, \eta_d(s^\text{Glyc}_2) = \frac{- 4\, \Delta_\text{ATP} G }{-\Delta_{\text{Glc}\to\text{Pyr}}  G}.
\end{equation}
From Eq.~\eqref{eq:Efficiency reverse transduction}, we also observe that a gear cannot be transducing for both $s$ and $s^{-1}$: If it does for the former, $0<\eta_{\mathfrak{g}}(s)<1$, it will be thermodynamically unfeasible for the latter, $\eta_{\mathfrak{g}}(s^{-1}) >1$.

\subsection{Conformal Gears}
\label{Sect:Conformal gears}
The concept of \emph{conformality}, introduced in~\cite{Muller_conformal_sums_Polyhedral_2016}, will play a key role in Sect.~\ref{Sect:Upper bound} for determining the optimal efficiency of a CRN engaged in transduction. Intuitively, conformality indicates whether a gear is utilized by the CRN---that is, whether it is `active' within the given stationary flux configuration. More precisely, a gear $\boldsymbol{\psi}_{\mathfrak{c}}$ is said to be \emph{conformal} to a flux $\boldsymbol{J}$ if it implements reactions in the same direction as $\boldsymbol{J}$: In mathematical terms, for every reaction $\rho$ where ${\psi}_{\mathfrak{c},\rho}\ne 0$, ${\psi}_{\mathfrak{c},\rho}$ and ${J}_\rho$ have the same sign. We emphasize that this notion of conformality depends exclusively on the direction of reaction fluxes, given by $\text{sign}(J_\rho)$, not on their magnitudes. 
In Fig.~\ref{fig:2}c, we show graphically the set of conformal gears for given fluxes in the GG and RF networks. If the direction of reaction fluxes is known only for a subset of reactions, as will be the case in Sect.~\ref{Sect:Applications}, the definition of conformal gears is restricted to this subset.
Finally, to avoid confusion since gears are defined up to a sign, a gear $\boldsymbol{\psi}_{\mathfrak{c}}$ is considered conformal as long as either $\boldsymbol{\psi}_{\mathfrak{c}}$ or $-\boldsymbol{\psi}_{\mathfrak{c}}$ aligns with $\boldsymbol{J}$.

\section{Optimal Efficiency}
\label{Sect:Optimal efficiency}
\begin{figure*}
    \centering
    \includegraphics[width=0.87\linewidth]{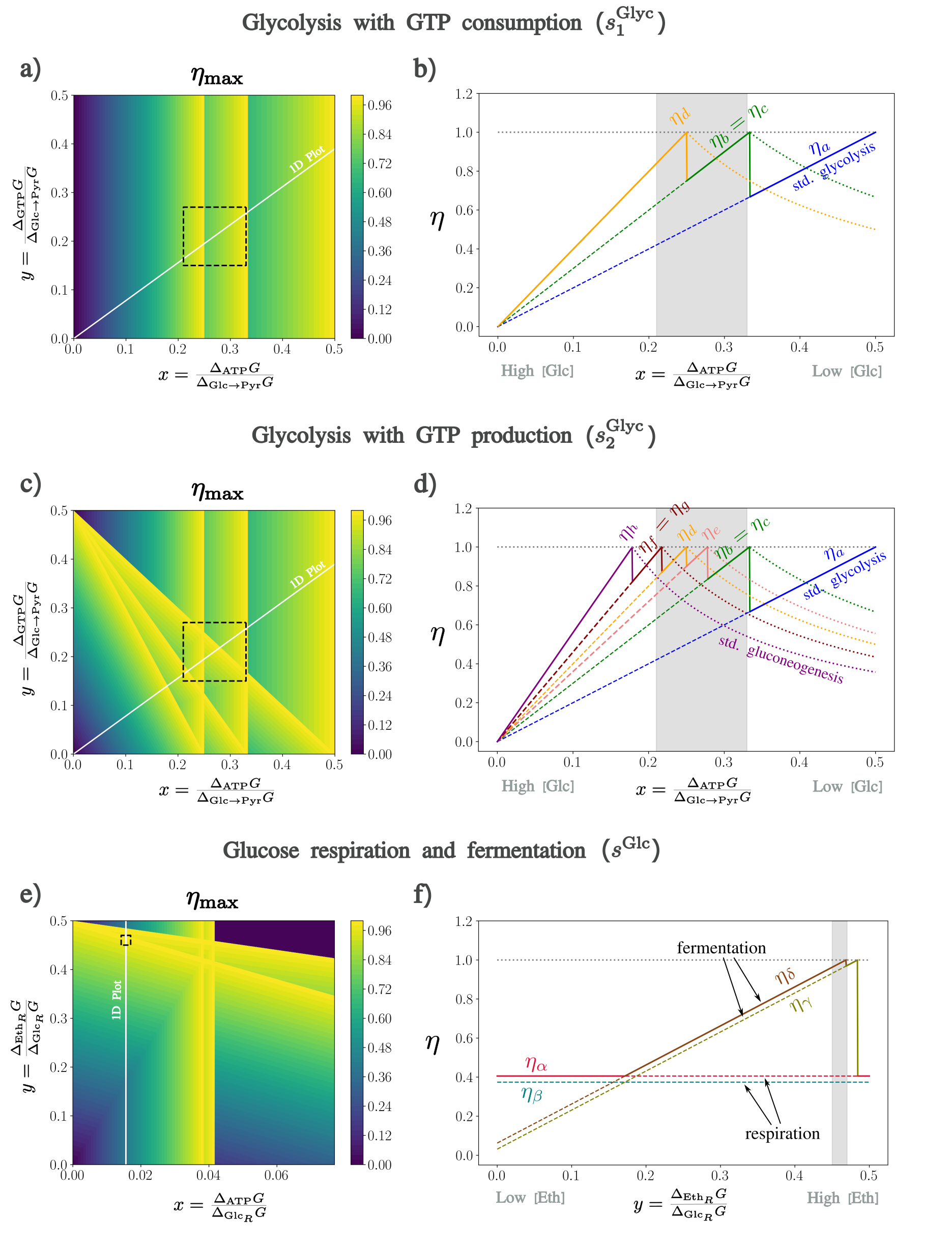}
    \caption{\textbf{Upper bound for the transduction efficiency.}
     \textbf{a) + b)} GG network performing glycolysis, while consuming GTP, $s^\text{Glyc}_1$ in Eq.~\eqref{eq:s Glycolisis 1}. \textbf{c) + d)} GG network performing glycolysis, while producing GTP, $s^\text{Glyc}_2$ in Eq.~\eqref{eq:s Glycolisis 2}. \textbf{e) + f)} RF network performing simultaneous glucose respiration and fermentation, $s^\text{Glc}$ in Eq.~\eqref{eq:s Glucose RF}.
    \textbf{Left:} Maximum transduction efficiency, determined by the optimal gear's efficiency, as a function of the operating conditions defined by $x$ and $y$. The black dashed square denotes the region corresponding to standard physiological conditions (Appendix~\ref{app:Standard physiological conditions}), reported in Table~\ref{Tab:Standard physiological conditions}.
    \textbf{Right:} 1D slice of operating conditions---corresponding to the white lines in the right panels. Dashed lines show suboptimal gear efficiencies; dotted lines show reverse transduction efficiencies where the respective gears are thermodynamically unfeasible in the forward direction.
    The shaded area denotes standard physiological conditions, analogously to the black dashed square on the right. Gears corresponding to the standard glycolytic and gluconeogenic pathways are labeled, as are those performing glucose respiration and fermentation.
    }
    \label{fig:3}
\end{figure*}

The second law of thermodynamics, as expressed in Eq.~\eqref{eq:Sigma decomposition_processes second law}, imposes only a broad constraint on efficiency $\eta < 1$ regardless of the system’s operating conditions. The key question is whether a more precise bound can be established by leveraging the topology of the CRN, without relying on any explicit information about reaction kinetics. Here, we show that even in the case of multi-process transduction, the second law can be refined at the level of individual gears. Unlike the single-input, single-output scenario~\cite{bilancioni2025gears}, this refinement yields multiple analytical bounds, each corresponding to a distinct input-output configuration.

In Sect.~\ref{Sect:Upper bound}, we present these refined bounds, along with a further refinement when the directionality of specific reaction fluxes is known. The proofs of these results are provided in Appendix~\ref{app:Upper bound}, and their application to the CRNs in Fig.~\ref{fig:1} is illustrated in Sect.~\ref{Sect:Applications}.

\subsection{Upper Bound for $\eta$}
\label{Sect:Upper bound}
For a CRN performing transduction  ($\eta > 0$), the second law refined at the gear-level imposes an upper bound on $\eta$ for each input-output split $s$: 
\begin{equation}
      \eta(s) \le \underset{\eta_\mathfrak{g}(s)<1}{\text{max}}  \,\,\eta_\mathfrak{g}(s).
      \label{eq:Upper bound general}
\end{equation}
Simply put, the most efficient thermodynamically feasible gear ($\eta_\mathfrak{g}(s)<1$) determines the bound. We call it the \emph{optimal gear}, and the bound is reached only if all the flux is concentrated on it. The intuition is similar to biking: to maximize the distance traveled per pedal stroke---the analog of efficiency---one must choose the heaviest gear that the terrain's steepness allows, with the steepness representing the operating conditions.
A direct implication of this inequality is that, for transduction to occur, the CRN must possess at least one transducing gear: $0< \eta_\mathfrak{g}(s)<1$. Importantly, this upper bound is derived solely from the CRN’s topology, without relying on any explicit information about the reaction fluxes. 

When the direction of certain reaction fluxes is known---for example, due to constraints imposed by the input-output split $s$, as will be the case in Sect.~\ref{Sect:Applications}---this upper bound can be further refined. Compared to Eq.~\eqref{eq:Upper bound general}, only the gears that are `active', \emph{i.e.}, conformal to the known flux directions (see Sect.~\ref{Sect:Conformal gears}), now appear on the right-hand side:
\begin{equation}
      \eta(s) \le \underset{\eta_\mathfrak{c}(s)<1}{\text{max}} \,\,\eta_\mathfrak{c}(s).
      \label{eq:Upper bound conformal}
\end{equation}


\subsection{Applications}
\label{Sect:Applications}

We now determine the optimal efficiency of the CRNs illustrated in Fig.~\ref{fig:1}, across  operating conditions. 
We first determine the constraints that the input-output configurations place on the directionality of the reaction fluxes within the CRN. 
We then identify the gears which are `active', \emph{i.e.}, conformal to these fluxes. This allows us to refine the upper bound using Eq.~\eqref{eq:Upper bound conformal} and find the optimal efficiency for any operating conditions. 
We end our analysis by comparing the predicted transducing gears and their efficiency with those measured in living cells.\\

\paragraph{GG Network.} We analyze the two input-output splits that correspond to glycolysis: with GTP consumption, ${s}^\text{Glyc}_1$ in Eq.~\eqref{eq:s Glycolisis 1}, and with GTP production, ${s}^\text{Glyc}_2$ in Eq.~\eqref{eq:s Glycolisis 2}. Starting from the former, Kirchhoff's law combined with the fact that GTP is hydrolyzed and glucose is converted to pyruvate implies
\begin{subequations}
    \label{eq:Constraints direction J Glyc_1}
    \begin{equation}
     J_{13}^\text{GG}, \, J_{14}^\text{GG} <0, 
    \end{equation}
    \begin{equation}
    J_{3}^\text{GG}, \, J_{6}^\text{GG},\,J_{7}^\text{GG}, \, J_{8}^\text{GG},\,J_{9}^\text{GG},\,J_{10}^\text{GG}, \, J_{11}^\text{GG},\,J_{12}^\text{GG}>0.
    \end{equation}
\end{subequations}
The sign convention of the reactions is given in Fig.~\ref{fig:1}a. These flux directions select $\boldsymbol{\psi}_a,\,\boldsymbol{\psi}_b,\,\boldsymbol{\psi}_c,\,\boldsymbol{\psi}_d,\,\boldsymbol{\psi}_i,\,\boldsymbol{\psi}_j,\,\boldsymbol{\psi}_k$ as conformal gears in Fig.~\ref{fig:2}a.
The last three exclusively hydrolyze either ATP or GTP and are therefore non-transducing, while the first four have efficiencies:
\begin{subequations}
\label{eq:Efficiency conformal gears Glyc_1}
    \begin{equation}
        \eta_a({s}^\text{Glyc}_1) =2 x,
    \end{equation} 
    \begin{equation}
       \eta_b({s}^\text{Glyc}_1) =\eta_c({s}^\text{Glyc}_1) =3 x,
    \end{equation} 
    \begin{equation}
        \eta_d({s}^\text{Glyc}_1) =4 x \;,
    \end{equation} 
\end{subequations}
 where we introduced the compact notation for the operating conditions:
\begin{equation}
    x =  \frac{ 
     \Delta_\text{ATP} G}{\Delta_{\text{Glc}\to\text{Pyr}}  G   }, \quad  \quad y = \frac{ 
     \Delta_\text{GTP} G}{\Delta_{\text{Glc}\to\text{Pyr}}  G   }.
\end{equation}
The upper bound in Eq.~\eqref{eq:Upper bound conformal} states that the most efficient gear among these four that is below one (the optimal gear) sets the maximum achievable efficiency. In Fig.~\ref{fig:3}a, we plot this upper bound as a function of the operating conditions $x$ and $y$. The black dashed square denotes the region corresponding to standard physiological conditions, defined in Appendix~\ref{app:Standard physiological conditions} and reported in Table~\ref{Tab:Standard physiological conditions}. In Fig.~\ref{fig:3}b, we report the optimal efficiency (full line) along a 1D line of operating conditions (given by the white line in Fig.~\ref{fig:3}a), as well as the efficiencies of suboptimal gears (dashed lines) and the reverse transduction efficiencies (dotted lines) in regions where the respective gears can only operate in the backward direction. Moving along the line of operating conditions can be thought of as varying glucose concentrations, and thus $\Delta_{\text{Glc}\to\text{Pyr}} G$, while keeping $\Delta_\text{ATP} G$ and $\Delta_\text{GTP} G$ fixed and equal to the values given in Table~\ref{Tab:Standard physiological conditions}. The shaded area, like the black dashed square in the 2D plot, denotes the physiological region.
We see that $\boldsymbol{\psi}_a,\,\boldsymbol{\psi}_b,\,\boldsymbol{\psi}_c$ are available to operate in the forward direction throughout this region, while $\boldsymbol{\psi}_d$ is available in the forward direction to the left and in the opposite direction to the right. We also observe that the gear corresponding to glycolysis, $\boldsymbol{\psi}_a$, has the lowest efficiency. 

One can repeat the same analysis for the input-output split ${s}^\text{Glyc}_2$, where GTP is synthesized instead of being hydrolyzed. This difference translates in the following constraints for the flux directions:
\begin{equation}
    J_{3}^\text{GG}, \, J_{6}^\text{GG},\,J_{7}^\text{GG}, \, J_{8}^\text{GG},\,J_{9}^\text{GG},\,J_{10}^\text{GG}, \, J_{11}^\text{GG},J_{13}^\text{GG}, \, J_{14}^\text{GG}>0. 
     \label{eq:Constraints direction J Glyc_1}
\end{equation}  
Now, all eleven gears in Fig.~\ref{fig:2}a are conformal to these flux directions. Among the eight gears that perform transduction, $\boldsymbol{\psi}_a,\,\boldsymbol{\psi}_b,\,\boldsymbol{\psi}_c,\,\boldsymbol{\psi}_d$ keep the same efficiency as in Eq.~\eqref{eq:Efficiency conformal gears Glyc_1} (since they do not involve GTP synthesis or hydrolysis), and $\boldsymbol{\psi}_e,\,\boldsymbol{\psi}_f,\,\boldsymbol{\psi}_g,\,\boldsymbol{\psi}_h$ have efficiencies: 
\begin{subequations}
    \label{eq:Efficiency conformal gears Glyc_2}
    \begin{equation}
        \eta_e({s}^\text{Glyc}_2) =2 x +2y,
    \end{equation} 
    \begin{equation}
       \eta_f({s}^\text{Glyc}_2) =\eta_g({s}^\text{Glyc}_2) =3 x+2y,
    \end{equation} 
    \begin{equation}
        \eta_h({s}^\text{Glyc}_2) =4 x+2y. 
    \end{equation} 
\end{subequations}
In Fig.~\ref{fig:3}c and \ref{fig:3}d, we report as before the upper bound of Eq.~\eqref{eq:Upper bound conformal}. Fig.~\ref{fig:3}d shows that both glycolysis and gluconeogenesis are thermodynamically feasible across the entire range of physiological conditions. For example,  glycolysis can be carried out by gear $\boldsymbol{\psi}_a$ since $\eta_a({s}^\text{Glyc}_2)<1$, while gluconeogenesis can be implemented by $-\boldsymbol{\psi}_h$, with $\eta_h({s}^\text{Gluc})=1/\eta_h({s}^\text{Glyc}_2)<1$ (dotted line). The simultaneous thermodynamic feasibility of glycolysis and gluconeogenesis is crucial as it confirms the known biochemical fact that enzyme regulation can control which of the two is activated under the same operating conditions. Indeed, in the human body during periods of low blood sugar, muscle cells engage in glycolysis while liver cells concurrently perform gluconeogenesis to maintain glucose homeostasis~\cite{Karimi_Fasting_impact_2021}. A toy model illustrating how enzymes can act as gear regulators can be found in ~\cite{bilancioni2025gears}.
Experimental evidence suggests that $\boldsymbol{\psi}_a$ and $\boldsymbol{\psi}_h$ are the only two gears utilized by living cells, aside from the non-transducing gears $\boldsymbol{\psi}_i$, $\boldsymbol{\psi}_j$, and $\boldsymbol{\psi}_k$.
Indeed, measurements of $\Delta_\rho G$ at the three branching points in Fig.~\ref{fig:1}a, yield for the upper reactions 1, 4, and 12~\cite{Park_Met_conc_FLux_Free_Energy_2016,Voet_Fundamentals_of_Biochemistry_2013}:
    \begin{equation}
        \Delta_{1} G,\,\Delta_{4} G,\,\Delta_{12} G <0 \implies J_{1}^\text{GG},\,J_{4}^\text{GG},\,J_{12}^\text{GG} >0,
    \end{equation}
and for the lower reactions 2, 5, 13, and 14:
    \begin{equation}
    \begin{split}
        &\Delta_{2} G,\,\Delta_{5} G,\,\Delta_{13} G,\,\Delta_{14} G>0 \\
       & \qquad \qquad \qquad \quad  \Downarrow\\
        &\qquad J_{2}^\text{GG},\,J_{5}^\text{GG},\,J_{13}^\text{GG},\,J_{14}^\text{GG} <0.
        \end{split}
    \end{equation}
The sign of $J_\rho$ follows from the local second law in Eq.~\eqref{eq:Local second law}. As a result, the only transducing gears that are conformal to these flux directions are $\boldsymbol{\psi}_a$ when the CRN performs glycolysis and $\boldsymbol{\psi}_h$ when the CRN performs gluconeogenesis, which explains why glycolysis in living cells is consistently associated with GTP consumption.
This observation raises an interesting question: Why do cells utilize only one transducing gear for each purpose, despite the thermodynamic feasibility of others (Fig.~\ref{fig:3}d)?
A possibility might be that, although $\eta_a$ and $\eta_h$
exhibit the lowest efficiencies for glycolysis and gluconeogenesis as shown in Fig.~\ref{fig:3}d, they are the only gears that remain available across the entire physiological region.
Arguments involving trade-offs between power and efficiency may also be at play, but assessing them requires kinetic knowledge.\\

\paragraph{RF Network.}
For this system, we analyze the case where yeast simultaneously performs glucose respiration and fermentation, ${s}^\text{Glc}$ in Eq.~\eqref{eq:s Glucose RF}, treating ethanol as a useful output. The fact that glucose is consumed while ethanol and CO$_2$ are produced places the following constraints on the reaction fluxes: 
\begin{equation}
    J_{1}^\text{RF}, \, J_{4}^\text{RF},\, J_{6}^\text{RF}>0.
\end{equation}
The sign convention of the reactions is given in Fig.~\ref{fig:1}c.
The gears conformal to these flux directions are $\boldsymbol{\psi}_\alpha,\,\boldsymbol{\psi}_\beta,\,\boldsymbol{\psi}_\gamma,\,\boldsymbol{\psi}_\delta,\,\boldsymbol{\psi}_\theta$ in Fig.~\ref{fig:2}b. 
While $\boldsymbol{\psi}_\theta$ exclusively hydrolyzes ATP and is, therefore, non-transducing, the others transduce with efficiencies:
\begin{subequations}
     \label{eq:Efficiency conformal gears Glc}
    \begin{equation}
        \eta_\alpha({s}^\text{Glc}) =(2n+2)\,x,
    \end{equation} 
    \begin{equation}
       \eta_\beta({s}^\text{Glc}) =2n\,x,
    \end{equation} 
    \begin{equation}
        \eta_\gamma({s}^\text{Glc}) =2x+ 2y, 
    \end{equation} 
    \begin{equation}
        \eta_\delta({s}^\text{Glc}) =4x+2y, 
    \end{equation} 
\end{subequations}
where $x$ and $y$ denote the operating conditions,
\begin{equation}
    x =  \frac{ 
     \Delta_\text{ATP} G}{\Delta_{\text{Glc}_R}  G}, \quad  \quad y = \frac{ 
     \Delta_{\text{Eth}_R} G}{\Delta_{\text{Glc}_R}  G},
\end{equation}
and $n$ is the number of ATP molecules synthesized by reaction 4 in Fig.~\ref{fig:1}c when an A-coA enters the TCA cycle and undergoes subsequent oxidative phosphorylation. This number $n$ is variable across living organisms and ranges from 8 to 18~\cite{Pfeiffer_Evol_perspect_Crabtree_2014}.

In Fig.~\ref{fig:3}e, we report the efficiencies of the gears as a function of the operating conditions $x$ and $y$ for $n =12$, keeping the same convention as before and with standard physiological conditions defined in Appendix~\ref{app:Standard physiological conditions} and summarized in Table~\ref{Tab:Standard physiological conditions}.
In Fig.~\ref{fig:3}f, the 1D slice of operating conditions is generated by varying ethanol concentrations, \emph{i.e.}, $\Delta_{\text{Eth}_R} G$, while keeping $\Delta_{\text{Glc}_R} G$ and $\Delta_\text{ATP} G$ fixed at the values specified in Table~\ref{Tab:Standard physiological conditions}. 
Two key findings emerge from this analysis.
First, unlike the GG network, yeast utilize three gears simultaneously. Indeed, measurements of metabolic fluxes in yeast undergoing concurrent respiration and fermentation under various experimental conditions yielded (see Fig. 3 in~\cite{Frick_13C_Flux_Yeast_metabolism_2005}):
\begin{equation}
    J_1^\text{RF},\, J_2^\text{RF},\,J_3^\text{RF},\,J_4^\text{RF},\,J_5^\text{RF},\,J_6^\text{RF}>0.
\end{equation}
The gears conformal to these flux directions are $\boldsymbol{\psi}_\alpha,\,\boldsymbol{\psi}_\beta,\,\boldsymbol{\psi}_\gamma$. 
Experimental evidence shows that shifting the operating conditions toward the lower left in Fig.~\ref{fig:3}e---\emph{i.e.}, increasing glucose concentration---drive yeast to upregulate fermentative metabolism (gear $\boldsymbol{\psi}_\gamma$), a phenomenon known as the Crabtree effect~\cite{Pfeiffer_Evol_perspect_Crabtree_2014,Crabtree_Effect_1966}.
Second, within the physiological range, fermentation---mediated by gears $\boldsymbol{\psi}_\gamma$ and $\boldsymbol{\psi}_\delta$---is more efficient than respiration, which relies on gears $\boldsymbol{\psi}_\alpha$ and $\boldsymbol{\psi}_\beta$ (when ethanol is considered as a useful output). Notably, fermentation achieves an efficiency exceeding 90\%, aligning with experimental observations~\cite{Teh_Thermo_analysis_yeast_fermentation_2010}. This finding underscores the importance of distinguishing thermodynamic efficiency from stoichiometric yield, a commonly used metric quantifying the number of ATP molecules produced per glucose molecule consumed. Although the stoichiometric yield of fermentation is significantly lower than that of respiration---since fermentation captures only a small fraction of glucose’s energy as ATP---its thermodynamic efficiency is higher.






 
\section{Discussion and Conclusions}
\label{Sec:DiscConc}

We significantly extended the traditional framework of energy transduction to encompass open CRNs with multiple resources. At its core, our approach is grounded in the concept of exergy and is formulated in terms of elementary processes, ensuring that all the energy transduction occurring within the CRN is captured.

We established a systematic yet versatile definition of thermodynamic efficiency that can be tailored based on what one considers useful outputs and what can be treated as immutable environmental variables.

By extending the concept of chemical gears to multi-process transduction, we showed that, even in such more complex settings, the network's topology constrains the optimal transduction efficiency achievable under given operating conditions, regardless of the kinetics.

Using our framework to analyze transduction in metabolic networks, we demonstrated its broad applicability and its ability to provide deep insights based on limited experimental data. 
We found that both glycolysis and gluconeogenesis operate in living cells using a single transducing gear, which corresponds to the least efficient one. In contrast, yeast simultaneously utilizes three transducing gears when performing respiration and fermentation. We highlighted the greater thermodynamic efficiency of fermentation over respiration when ethanol is considered a useful output, despite the fact that the opposite conclusion is reached when comparing the stoichiometric yields of these pathways.
Furthermore, we characterized the optimal efficiency of these metabolic networks under any operating condition. Our results call for future experimental studies which could leverage these optimal values to better understand the extent to which metabolic networks are shaped by the need to maintain high thermodynamic efficiencies.

We have seen that, although our optimal efficiency predictions are solely based on the network's topology and the operating conditions, additional constraints can be seamlessly incorporated into the framework. Specifically, constraints on flux directionality can be used to restrict the set of possible conformal gears in Eq.~\eqref{eq:Upper bound conformal}, further refining the efficiency bound. Such constraints may be derived from: experimental information about some fluxes in the CRN, concentrations and standard Gibbs free energies of some internal species~\cite{Kummel_NET_analysis_2006, HENRY_TMFA_2007,HoppeIncluding_metabolite_concentrations2007,ZambonianNET_software2008,Jol_System_Level_Insights_east_Metabolism_2012,HARALDSDOTTIR_Assignment_Reaction_Directionality2012,NoorEstimation_Gibbs_2013,ATAMAN_thermodynamics_based_network_analysi2015,Heirendt_COBRA_2019}, transcriptional regulation~\cite{JUNGREUTHMAYER_Constraint_transcriptional_regulatory2013}, or kinetic modeling~\cite{Yasemi_Constraint_Based_review_2021}.

We developed our transduction theory focusing on chemical free energy under isothermal and well-mixed conditions. However, the framework is more general and can accommodate other free energy sources, such as thermal gradients in nonisothermal or photochemical systems~\cite{penocchioNonequilibriumThermodynamicsLightinduced2021}, voltage differences in electrochemical systems, and concentration gradients in compartmentalized systems~\cite{avanziniThermodynamicsChemicalWaves2019,Avanzini_React_diff_implications2024}.
These extensions should not constitute major conceptual obstacles, as the fundamental approach---based on establishing a reference equilibrium and identifying elementary processes---remains unchanged. The only difference involves including additional elementary processes associated with energy or electron transfer between reservoirs at different temperatures or voltages, and treating species in separate compartments as distinct.
We also focused on steady-state transduction, but incorporating changes occurring in the chemostats' free energy is a natural next step.

We finally note that, since our analysis of transduction is systematic, it can be implemented algorithmically in computer code and used to efficiently analyze complex CRNs. The only required inputs are the CRN stoichiometric matrix $\mathbb{S}$, the selected chemostatted species $Y$, their chemical potentials, and the atomic composition matrix $\mathbb{A}^Y$ for these species.

\section*{Acknowledgements} 
MB is funded by AFR PhD grant 15749869 funded by Luxembourg National Research Fund (FNR) and ME by CORE project ChemComplex (Grant No. C21/MS/16356329), and by project INTER/FNRS/20/15074473 funded by F.R.S.-FNRS (Belgium) and FNR.\\

\section{Author contributions}

MB and ME contributed equally to all aspects of this
work, including the conceptualization, design, analysis,
and writing of the manuscript.

\section{Conflict of interest}
The authors declare that they have no conflicts to disclose.

\section{Data Availability}
The data that support the findings of this study are available within the article.



\appendix
\section{Elementary Processes}
\label{app:Elementary processes}
A process $\boldsymbol{p}$ is \emph{elementary} if it involves a minimal set of species  $Y$. Specifically, $\boldsymbol{p}$ is said to involve a species $y$ if $p_y \neq 0$, and the set of species is minimal if removing any one of them would make it impossible to construct a valid process with the remaining species.
From this definition, it follows that an elementary process $\boldsymbol{p}$ cannot have any subprocess.
Formally, $\boldsymbol{p}'$ is a subprocess of $\boldsymbol{p}$ if $|p_y| \geq |p'_y|$ and $p_y, p'_y \geq 0$ for all $y$. That is, the reactants of $\boldsymbol{p}'$ are a subset of the reactants of $\boldsymbol{p}$ and same for the products.
If $\boldsymbol{p}$ were elementary but had a subprocess $\boldsymbol{p}'$, one could construct a new process $\boldsymbol{p} - \lambda \boldsymbol{p}'$ (for a suitable $\lambda$), which would involve fewer species than $\boldsymbol{p}$. This would contradict the minimality condition of $\boldsymbol{p}$.

We emphasize that the notion of elementary processes depends solely on the composition of the $Y$ species and not on the CRN. In particular, it is completely unrelated to the number of reaction steps needed to realize an elementary process.

The mathematical criterion for determining whether a process $\boldsymbol{p}$ is elementary is 
\begin{equation}
    \text{dim(ker(}\mathbb{A}^Y_{\boldsymbol{p}}\text{))} = 1,
\end{equation}
where $\mathbb{A}^Y_{\boldsymbol{p}}$ is the atomic composition matrix reduced to the $Y$ species for which $p_y \neq 0$. Since $\mathbb{A}^Y_{\boldsymbol{p}}\boldsymbol{p} = 0$, it follows that the kernel of $\mathbb{A}^Y_{\boldsymbol{p}}$ has at least dimension 1. If $\text{dim(ker(}\mathbb{A}^Y_{\boldsymbol{p}}\text{))} > 1$, there would exist two independent processes $\boldsymbol{p}'$ and $\boldsymbol{p}''$ in Ker$(\mathbb{A}^Y_{\boldsymbol{p}})$. Linear combinations of these processes could then generate a new process involving fewer species than $\boldsymbol{p}$, contradicting the minimality condition.
Through this criterion, it is possible to verify that all processes in Eq.~\eqref{eq:All GG RF processes} are elementary.
We underscore that the definition of elementary processes is mathematically equivalent to that of elementary flux modes, as defined in Sect.~\ref{Sect:EFMs}. Consequently, existing algorithms for identifying the latter~\cite{Terzer_efmtool_2008} are directly applicable to the former. 

\subsection{Existence of an Elementary Subprocess} 
\label{App:Existence of an elementary subprocess}
If a process $\boldsymbol{p}$ is non-elementary, one can always find an elementary subprocess within $\boldsymbol{p}$.
The proof proceeds as follows:
By definition, there exists a process $\boldsymbol{p}'$ that involves a smaller set of $Y$ species.
If, for  any $\delta$,  $\delta\,\boldsymbol{p}'$ is a subprocess of $\boldsymbol{p}$,  we set  $\boldsymbol{p}'' =\delta\,\boldsymbol{p}'$. If not, we construct $\boldsymbol{p}'' =\delta\,(\boldsymbol{p} -\lambda \boldsymbol{p}')$ where $\lambda >0 $ is the smallest value for which $\lambda \boldsymbol{p}'$ cancels a component of $\boldsymbol{p}$ and $\delta$ a proper rescaling factor that makes $\boldsymbol{p}''$ a subprocess of $\boldsymbol{p}$. If the resulting subprocess $\boldsymbol{p}''$ is elementary, the procedure terminates. Otherwise, the above steps are repeated iteratively until an elementary subprocess is obtained.

\section{Characterization of $\Pi_\text{CRN}$}
\label{app:Characterization of PI_CRN}
In this appendix, we show that if ${\boldsymbol{\phi}_\epsilon}$ is a chosen set of emergent cycles, then the corresponding implemented processes $\boldsymbol{p}_\epsilon = \mathbb{S}^Y \boldsymbol{\phi}_\epsilon$ form a basis for $\Pi_\text{CRN}$.
$\Pi_\text{CRN}$, as defined in Eq.~\eqref{eq:Processes executable by CRN}, represents the subset of processes that a CRN can execute. Since, from Sect.~\ref{Sect:Cycles}, $\boldsymbol{J}$ is a cycle, a process $\boldsymbol{p}$ belongs to $\Pi_\text{CRN}$ if and only if there exists a cycle $\boldsymbol{\psi}$ such that 
$\boldsymbol{p}=\mathbb{S}^Y\boldsymbol{\psi}$.
As a first step, we prove that $\{\boldsymbol{p}_\epsilon\}$ is a linearly independent set.
Suppose, for contradiction, that it is not. Then, there would exist a non-trivial linear combination such that:
\begin{equation}
    \sum_\epsilon a_\epsilon \boldsymbol{p}_\epsilon = 0.
\end{equation}
Substituting $\boldsymbol{p}_\epsilon = \mathbb{S}^Y \boldsymbol{\phi}_\epsilon$, we would have:
\begin{equation}
   \sum_\epsilon a_\epsilon \boldsymbol{p}_\epsilon = 0 \implies \mathbb{S}^Y\left(\sum_\epsilon a_\epsilon \boldsymbol{\phi}_\epsilon\right) = 0,
\end{equation}
implying that $\sum_\epsilon a_\epsilon \boldsymbol{\phi}_\epsilon$ is an internal cycle. However, this would contradict the assumption that $\{\boldsymbol{\phi}_i\}\cup\{\boldsymbol{\phi}_\epsilon\}$ is a basis. 
Next, we show that $\{\boldsymbol{p}_\epsilon\}$ is a complete set, meaning it can represent any process in $\Pi_\text{CRN}$. By using the fact that any cycle $\boldsymbol{\psi}$ can be written in terms of the basis $\{\boldsymbol{\phi}_i\}\cup\{\boldsymbol{\phi}_\epsilon\}$, one has
\begin{equation}
    \boldsymbol{p}=\mathbb{S}^Y\boldsymbol{\psi} = \mathbb{S}^Y\left(\sum_i a_i \boldsymbol{\phi}_i+  \sum_\epsilon a_\epsilon \boldsymbol{\phi}_\epsilon\right) =  \sum_\epsilon a_\epsilon  \boldsymbol{p}_\epsilon.
\end{equation}

\section{Processes Associated with the Force Species}
\label{App:Processes associated with Y_F}

In this section, we demonstrate that for each species $y_F$ there exists a process $\boldsymbol{p}_F = \mathbb{S}^Y\boldsymbol{\psi}_F$ that involves only $y_F$ and the potential species $Y_P$~\cite{raoConservationLawsWork2018}.

First of all, we note that, 
for Eq.~\eqref{eq: Equilibrium condition Delta G} to hold, no process in $\Pi_\text{CRN}$ can involve only $Y_P$ species or, equivalently, there must be no external cycle that leaves all $Y_F$ species unchanged.
Mathematically, let the matrix $\mathbb{S}^{X + Y_F}$ be defined as
\begin{equation}
   \mathbb{S}^{X + Y_F} =  \begin{pmatrix} \mathbb{S}^X \\ \mathbb{S}^{Y_F} \end{pmatrix},
   \label{eq:Matrix S(X+Y_F)}
\end{equation}
where $\mathbb{S}^{Y_F}$ is the stoichiometric matrix reduced to species $Y_F$.
The above requirement translates into the fact that Ker($\mathbb{S}^{X + Y_F} $)---the subspace of cycles that leave both $X$ and $Y_F$ species unchanged---must contain only internal cycles. This is satisfied if
\begin{equation}
    \text{dim(ker(}\mathbb{S}^{X + Y_F} )) =\text{dim(ker(}\mathbb{S} ))  =|i|,
    \label{eq:Condition for valid partitioning Y}
\end{equation}
where $|i|$ is the number of independent internal cycles.
Now, we analyze the matrix $\mathbb{S}^{X + Y_F-y_F}$, which is obtained from $\mathbb{S}^{X + Y_F}$, in Eq.~\eqref{eq:Matrix S(X+Y_F)},  by removing the row corresponding to species $y_F$. Our goal is to show that the kernel of this reduced matrix contains exactly one emergent cycle.
Using Eq.~\eqref{eq:Condition for valid partitioning Y}, we note that:
\begin{equation}
    \text{dim(ker(}\mathbb{S}^{X} )) - \text{dim(ker(}\mathbb{S}^{X + Y_F} ))  = |\epsilon| =|Y_F|.
\end{equation}
This change in kernel dimensionality upon removing the  $|Y_F|$ rows corresponding to the force species from $\mathbb{S}^{X + Y_F}$ indicates that these rows impose $|Y_F|$  independent constraints.
Therefore,
removing any of these rows increases the dimension of the kernel by exactly one. Consequently, the matrix $\mathbb{S}^{X + Y_F-y_F}$ acquires one additional independent cycle in its kernel, which must necessarily be external  since all internal cycles are already present in the kernel. Let $\boldsymbol{\psi}_F$ denote such a cycle. Since 
\begin{equation}
    \mathbb{S}^{X + Y_F-y_F}\boldsymbol{\psi}_F = 0,
\end{equation}
the corresponding process, $\boldsymbol{p}_F = \mathbb{S}^Y\boldsymbol{\psi}_F$, only involves $y_F$ and species $Y_P$. 
Since each $\boldsymbol{p}_F$ involves a different  $y_F$, the set $\{\boldsymbol{\psi}_F\}$ is linearly independent. Given that $|Y_F| = |\epsilon|$, this set can be taken as the set of emergent cycles. Moreover, following Appendix~\ref{app:Characterization of PI_CRN}, the set $\{\boldsymbol{p}_F\}$ forms a basis for $\Pi_\text{CRN}$ and is, therefore, complete. We note that, although the selection of $\boldsymbol{\psi}_F$ is not unique since linear combinations of $\boldsymbol{\psi}_F$ with internal cycles are also valid choices, the resulting $\boldsymbol{p}_F$ remains uniquely determined.

To provide an example, consider the partitions $Y_F^\text{GG} = \{\text{Glc},\text{GTP},\text{ATP} \}$ and $Y_F^\text{RF} = \{\text{Glc},\text{Eth},\text{ATP} \}$ used throughout this paper. In the two cases, ATP identifies as two possible emergent cycles: 
\begin{equation}
\boldsymbol{\psi}_{\text{ATP}}^\text{GG}  =
\,
\begin{array}{l}
\scriptstyle \textcolor{gray}{1}\\
\scriptstyle \textcolor{gray}{2}
\end{array}
\begin{pmatrix}
+1\\
-1 \\
\end{pmatrix},\quad
\boldsymbol{\psi}_\text{ATP}^\text{RF} = 
\, 
\begin{array}{l}
\scriptstyle \textcolor{gray}{2}\\
\scriptstyle \textcolor{gray}{3}\\
\scriptstyle \textcolor{gray}{5}
\end{array}
\begin{pmatrix}
-1\\
+1 \\
+1
\end{pmatrix},
\end{equation}
and the process implemented by these emergent cycles is $\boldsymbol{p}_\text{ATP}$, in Eq.~\eqref{eq:All GG RF processes}.

\section{Upper Bound for $\eta$}
\label{app:Upper bound}

In this section, we prove Eqs.~\eqref{eq:Upper bound general} and \eqref{eq:Upper bound conformal}. 
As a key intermediate step, in Sect.~\ref{app:Conformal decomposition of the entropy production}, we first revisit the conformal decomposition of the entropy production, which was derived in our previous work (see SI~\cite{bilancioni2025gears}).

\subsection{Conformal Decomposition of the Entropy Production}
\label{app:Conformal decomposition of the entropy production}
The stationary flux and  entropy production can be non-uniquely decomposed in terms of conformal gears~\cite{bilancioni2025gears}:
\begin{subequations}
    \begin{equation}
        \boldsymbol{J} = \sum_\mathfrak{c} j_\mathfrak{c} \boldsymbol{\psi}_\mathfrak{c},
    \end{equation}
    \begin{equation}
        \dot\Sigma  = \sum_\mathfrak{c} - j_\mathfrak{c} \Delta_\mathfrak{c}  G>0,
    \label{eq:conformal decomposition EP}
    \end{equation}
\end{subequations}
where $\Delta_\mathfrak{c}  G = \boldsymbol{\mu}\cdot\boldsymbol{\psi}_c $ represents the Gibbs free energy change associated with gear $\boldsymbol{\psi}_c$.
The key properties of these decompositions are that, for each $\mathfrak{c}$, $j_\mathfrak{c} \boldsymbol{\psi}_\mathfrak{c} $ is conformal to $\boldsymbol{J}$ and 
\begin{equation}
    -j_\mathfrak{c}\Delta_\mathfrak{c} G>0,
    \label{eq:each c contributes positively to EP}
\end{equation}
meaning that each gear contributes positively to entropy production. By using
 \begin{equation}
     \Delta_\mathfrak{c}  G = \Delta_\mathfrak{c}  G_\text{in}(s) + \Delta_\mathfrak{c}  G_\text{out}(s),
 \end{equation}
from Eq.~\eqref{eq:Gear delta G decomposition_processes split}, the input and output parts of the entropy production,
  \begin{equation}
     \dot\Sigma = \dot\Sigma_\text{in}(s) + \dot\Sigma_\text{out}(s), 
 \end{equation}
can be expressed as
 \begin{subequations}
     \label{eq:conformal decomposition i-o parts of EP}
 \begin{equation}
    \dot\Sigma_\text{in}(s) = \sum_\mathfrak{c} - j_\mathfrak{c} \Delta_\mathfrak{c}  G_\text{in}(s),
    \end{equation}
    \begin{equation}
    \dot\Sigma_\text{out}(s) = \sum_\mathfrak{c} - j_\mathfrak{c} \Delta_\mathfrak{c}  G_\text{out}(s).
    \end{equation}
 \end{subequations}
The decomposition in Eq.~\eqref{eq:conformal decomposition EP} relies on the local validity of the second law, explained in Sect.~\ref{Sect:Local validity of the Second law}.

\subsection{Proof of the Upper Bound}
\label{subapp: Proof of the upper bound}

We prove the refined upper bound of Eq.~\eqref{eq:Upper bound conformal} for a CRN performing transduction, $0<\eta(s)<1$. From this,  Eq.~\eqref{eq:Upper bound general} automatically follows. Along the proof, it becomes apparent that all non-transducing gears negatively affect efficiency.

We start by rewriting the efficiency in Eq.~\eqref{eq:Definition efficiency} using Eq.~\eqref{eq:conformal decomposition i-o parts of EP}:
\begin{equation}
     \eta(s) = \frac{-\dot\Sigma_\text{out}(s) }{\dot\Sigma_\text{in}(s) } = \frac{\sum_\mathfrak{c} j_\mathfrak{c} \Delta_\mathfrak{c} G_\text{out}(s) }{-\sum_\mathfrak{c} j_\mathfrak{c} \Delta_\mathfrak{c} G_\text{in}(s) }, 
\end{equation} 
Gears for which $\Delta_\mathfrak{c} G_\text{in}(s) =0$ have a detrimental effect on the transduction efficiency since they contribute solely to the numerator with the term $j_\mathfrak{c} \Delta_\mathfrak{c} G_\text{out}(s)$, which is negative according to Eq.~\eqref{eq:each c contributes positively to EP}. In the following, we restrict to the case where $\Delta_\mathfrak{c} G_\text{in}(s) \ne 0$ for all gears and we define the forward direction as the one having $- \Delta_\mathfrak{c} G_\text{in}(s) >0$. 
Using the definition of gear efficiency, 
\begin{equation}
    \eta_\mathfrak{c}(s)  = \frac{ \Delta_\mathfrak{c} G_\text{out}(s) }{-\Delta_\mathfrak{c} G_\text{in}(s) }, 
\end{equation}
the transduction efficiency can be rewritten as 
\begin{equation}
     \eta(s)  =   \frac{-\sum_\mathfrak{c} j_\mathfrak{c} \Delta_\mathfrak{c} G_\text{in}(s)\, \eta_\mathfrak{c} }{ -\sum_\mathfrak{c} j_\mathfrak{c} \Delta_\mathfrak{c} G_\text{in}(s) }. 
     \label{eq: efficiency in terms of conformal gear efficiencies}
\end{equation} 
Since $-j_\mathfrak{c}\Delta_\mathfrak{c} G>0$ from Eq.~\eqref{eq:each c contributes positively to EP}, the following relation holds between $\eta_\mathfrak{c}(s)$ and $j_\mathfrak{c}$: 
\begin{subequations}
    \begin{equation}
        \eta_\mathfrak{c}(s) < 1 \iff \Delta_\mathfrak{c} G <0 \iff j_\mathfrak{c} > 0,
    \end{equation}
    \begin{equation}
        \eta_\mathfrak{c}(s) > 1\iff \Delta_\mathfrak{c} G > 0  \iff j_\mathfrak{c} < 0.
    \end{equation}
\end{subequations}
Based on this, we split the set of conformal gears into thermodynamically feasible and unfeasible ones: $ \{\mathfrak{c}\} = \{\mathfrak{c}'\} + \{\mathfrak{c}''\}$, with $\eta_\mathfrak{c'}(s)< 1$ and $\eta_\mathfrak{c''}(s)> 1$. One has $j_\mathfrak{c'}  > 0$ and $ j_\mathfrak{c''}  < 0$, which implies that the coefficients $r_\mathfrak{c'} = -j_\mathfrak{c'} \Delta_\mathfrak{c'} G_\text{in}(s)$ and $q_\mathfrak{c''} = j_\mathfrak{c''} \Delta_\mathfrak{c''} G_\text{in}(s)$ are both positive. We can express the efficiency in Eq.~\eqref{eq: efficiency in terms of conformal gear efficiencies} as 
\begin{equation}
      \eta(s) = \frac{\sum_\mathfrak{c'}  r_\mathfrak{c'}\eta_\mathfrak{c'}(s)  - \sum_\mathfrak{c''}  q_\mathfrak{c''}\eta_\mathfrak{c''}(s)}{\sum_\mathfrak{c'}  r_\mathfrak{c'}-\sum_\mathfrak{c''}  q_\mathfrak{c''}}.
\end{equation}
 We first show that every $q_\mathfrak{c''}\ne 0$ impacts negatively the efficiency compared to the case where the same $q_\mathfrak{c''}$ is zero. 
To do this, we rewrite the efficiency as follows:
\begin{equation}
      \eta(s) = \frac{ C -   q_\mathfrak{c''}\eta_\mathfrak{c''}(s)}{D-  q_\mathfrak{c''}},
\end{equation}
where all the other terms have been reabsorbed into the constants $C$ and $D$. Since the CRN is transducing, $\dot\Sigma_\text{in}(s) >0$  and $\dot\Sigma_\text{out}(s)<0$, both numerator and denominator are positive. In addition, $\eta(s)<1$ implies  $C<D\,\eta_\mathfrak{c''}(s)$, which is enough to prove that $\eta(s)<C/D.$ Repeating this argument for all the conformal gears in $ \{c''\}$, one obtains:
\begin{equation}
      \eta(s) \le \frac{\sum_\mathfrak{c'}  r_\mathfrak{c'}\eta_\mathfrak{c'}(s)}{\sum_\mathfrak{c'}  r_\mathfrak{c'}}.
\end{equation}
Now, the RHS is simply a weighted average with positive coefficients and, therefore, we obtain the desired inequality:
\begin{equation}
      \eta(s) \le \underset{\eta_\mathfrak{c}(s)<1}{\text{max}} \,\,\eta_\mathfrak{c}(s).
      \label{eq: Appendix Upper bound conformal}
\end{equation}

\section{Standard Physiological Conditions} 
\label{app:Standard physiological conditions}
Table~\ref{Tab:Standard physiological conditions} lists the Gibbs free energy changes of all elementary processes in Eq.~\eqref{eq:All GG RF processes}. They reflect typical physiological values, referred to as `standard physiological conditions'~\cite{wachtelFreeenergyTransductionChemical2022}, and are sourced from eQuilibrator~\cite{Beber_Equilibrator_2021}.

Specifically, under these conditions: pH
= 7.5, pMg = 3.0, I = 0.25 (ionic strength), T = 298.15 K. All metabolite concentrations are uniformly set to 1 millimolar, the typical order of magnitude of physiological concentrations, with the exception of water, where [H$_2$O] = 55 M.
\begin{table}[h!]
\centering
\caption{Processes from Eq.~\eqref{eq:All GG RF processes} and their corresponding free energy changes under standard physiological conditions from~\cite{Beber_Equilibrator_2021}.}
\begin{tabular}{@{}lc@{}}
\toprule
\textbf{Chemical Process}   & \textbf{$-\Delta G$ (kJ mol$^{-1}$)}  \\ \midrule
Glucose $\rightarrow$ Pyruvate          & $170 \pm 40$  ($= 68 \pm 15 RT$)          \\
GTP Hydrolysis              & $35 \pm 7$ ($= 14 \pm 3 RT$)           \\
ATP Hydrolysis              & $46 \pm 4$  ($= 18.4 \pm 1.7 RT$)    \\
Glucose Respiration         & $2910 \pm 50$ ($= 1180 \pm 20 RT$)             \\
Ethanol Respiration         & $1350 \pm 25$ ($= 540 \pm 10 RT$)        \\ \bottomrule
\end{tabular}
\label{Tab:Standard physiological conditions}
\end{table}

\end{document}